\input epsf.sty 
\documentclass[epsf]{aa} 
\def\Ess{E^{\rm (ss)}}
\def\Esd{E^{\rm (sd)}}
\def\ief{i_{\rm ef}}
\def\MBH{M_{\rm BH}}
\def\Ms{M_{\rm s}}
\def\ms{m_{\rm s}}
\def\ns{n_{\rm s}}
\def\ndoty{\dot n _{\rm y}}
\def\qm2{q_{\rm m-2}}
\def\Rg{R_{\rm g}}
\def\Rs{R_{\rm s}}
\def\rs{r_{\rm s}}
\def\Sigd{\Sigma_{\rm d}}
\def\Td{T_{\rm d}}
\def\Vs{V_{\rm s}}
\def\Vsr{V_{\rm sr}}
\def\VR{V_{\rm R}}

\begin{document} 
 \thesaurus{03           
              (02.18.7;  
               11.01.2;  
	       11.19.1;  
               13.21.1;  
               13.25.2)}  
\title{The role of the central stellar cluster in active galactic nuclei.}  
\subtitle{I. Semi-analytical model}  
 
\author{E. Y. Vilkoviskij 
\inst{1,2} 
\and  B. Czerny
\inst{3} 
} 
 
\offprints{E. Vilkoviskij}    
 
\institute{  
Fesenkov Astrophysical Institute, Observatory, 480020 
Almaty, Kazakstan \\  email: vilk@afi.south-capital.kz 
\and 
Isaac Newton Institute of Chile, the Kazakhstan 
branch 
\and Copernicus Astronomical Center, Bartycka 18, 00-716 Warsaw, Poland\\ 
email: bcz@camk.edu.pl} 
\date{Received ...; accepted ...} 
 
\maketitle 

\begin{abstract}
The subject of the paper is the role of the massive stellar cluster in the
activity phenomenon and in the structure of active galactic nuclei. 
We introduce a simple model
of stellar dynamics in the internal part of the cluster, which allows us
to include both the star-disk and the star-star interactions. It is shown
that the properties of the distribution of stars in the vicinity of the
black hole are determined both by the interaction of the stars with the
accretion disk and by the pair gravitational and contact interaction between
the stars. We calculate the distribution of stars in the central parts of
the cluster and we discuss possible effects of stellar mass-loss due to the
star-disk interaction. Finally, we study the implications of the central
cluster for active galactic nuclei activity. We model the broad line 
region assuming that the
gaseous wakes, following stars after each disk crossing, play the role of
the broad line region clouds, and we calculate the corresponding line
profiles. We also analyze the contribution of star-star and star-disk
collisions to active galactic nuclei variability.

\end{abstract}

\thesaurus{03                         (02.18.7;                 11.01.2;               11.19.1;                 13.21.1;                 13.25.2)}
%
\section{ Introduction}

The idea of an important role of the compact stellar cluster (hereafter CSC)
in creation and evolution of active galactic nuclei (hereafter AGN) 
was formulated at the very early stage of
AGN investigation (e.g., Shklovskiy 1964), and the evolution of a gas sphere
containing a CSC was first considered in the work of Unno (1971). The
dissipative interaction of the compact stellar cluster with the massive
central object in the form of a super-star was considered by Vilkoviskij
(1975) and Hara (1978), and the tidal interaction of stars with the central
object in the form of a massive black hole (hereafter MBH) - by Hills (1975). 
The
evolution of the dense, non-rotating stellar cluster was discussed by Saslaw
(1966), Bisnovatyi-Kogan (1978), and others. The interaction of stars with
the accretion disk (hereafter AD) 
was considered in the works by Vilkoviskij and
Bekbosarov (1981), Vilkoviskij (1983), Zentsova (1983), Syer et al. (1991). 
More detailed investigations of the evolution of the stellar
orbits, crossing the AD were presented in Artymowicz et al., (1993), Karas
and Vokrouhlicky (1993), Rauch (1995), Vokrouhlicky \& Karas (1993, 1998),
Subr \& Karas (1999), and others. The role of the outer parts of AD (which
can be gravitationally unstable) in the global evolution of AGNs through
efficient star formation was discussed by Collin \& Zahn (1999).

A general investigation of the problems of CSC structure and evolution,
taking into account the interaction of the star cluster with the accretion
disk, demands calculations of very complicated numerical models. In
contrast, the main aim of the present paper is a preliminary analysis at a
qualitative level, but including both star-star and star-disk interactions.
The star-disk and the star-star interactions were not considered before as
acting simultaneously. Rauch (1995, 1999) considered and numerically
calculated the cases when either star-disk or star-star collisions dominate;
we consider here a model which is very simple but includes the joint action
of the star-disk and star-star interactions. This permits us to find the
space distribution of the stars around ADs. As an application of this
result, we consider star-related model of Broad Emission Line Regions 
(hereafter BELRs) and the possible role of the
inelastic interactions of stars both with the disk and other stars in the
variability of AGNs.

The plan of the paper is as follows. In Sect.~2 we consider the stellar
dynamics in the central part of CSC in AGNs. In Sect.~3 we derive the
structure of the stellar cluster in the AD region, including the role of
contact stellar collisions and some effects of the mass loss of the stars.
The application of the results to modeling BELR and variability are discussed
in Sects.~4 and 5, correspondingly. The discussion of the results and
conclusions are given in Sect.~6.

\section{Stellar dynamics in the central part of CSC}

\subsection{The star-disk interaction}

We assume that there are massive and compact stellar clusters in the centers
of AGNs. Observations of the nearest galaxies show clearly that dense
stellar clusters do exist in their centers, with sizes close to a parsec,
with masses $10^7-10^8~ M_{\odot }$, and central stellar densities $
10^6-10^7 ~{\rm pc}^{-3}$. It is reasonable to suppose that in AGNs the
density in CSCs is increased, and in the case of the more powerful AGNs the
central stellar density can reach values of up to $10^8-10^9~{\rm pc}^{-3}$,
as follows in particular from the gas-dynamic solutions for the broad
absorption line QSOs (hereafter 
BAL QSO, see Vilkoviskij et al. 1999). In contrast to
the case of close galaxies with low activity, the observational data about
CSCs in luminous AGNs are very poor; we will assume that those CSCs
are massive, rotate, and consist predominantly of low-mass stars.

For the sake of qualitative analysis, we introduce several
simplifications in our consideration: (1) we consider equal stellar masses,
(2) we assume that all stellar orbits are almost circular, which is a good
approximation in the internal part of the CSC only (see below), (3) we
introduce some effective values of the main physical parameters -- the
inclination angle of stellar orbits to the plane of AD, and the stellar
density (see the next subsection), and we investigate the radial dependence
of these parameters.

The star-disk (hereafter also s-d) interactions were considered in a
series of studies. The main, and quite obvious result of s-d interaction is
that the orbits of stars that have been captured into bound orbits
crossing the accretion disk evolve from eccentric to almost circular, 
with diminishing radius and inclination angle
(e.g., Syer et al. 1991). The planes of the orbits approach the
plane of the 
AD in the process of the orbit evolution. As a result, the stars on
these evolved orbits move mostly in the direction of the AD rotation, due to
the dissipative interaction of the stars with the disk. For a single star,
this process leads finally to the incorporation of the stellar orbit
into the disk plane, and the further fate of such stars is undefined
(Artymowicz et al. 1993). In the next sections we will show that
consideration of the {\it collective} star inflow through the disk instead
of the {\it single} stellar orbit evolution change the result: due to high
stellar density, the star-star interaction scatters stellar orbits,
thus counteracting the capture of stars by the disk. This effect allows the
CSC to reach some equilibrium, with fluctuating but non-vanishing
inclination angle of the orbits, and the dissipative inflow of the stars
continues to the center of the AD.

In the present section we will shortly summarize the properties of the
individual star-disk interaction. For simplicity we will take into account
two above-mentioned basic theoretically deduced properties of the orbits,
crossing the disk: we assume that all orbits are close to the circular
shape, and that the inclination angles of the orbits to the disk are small.

Let us consider a star in an orbit, crossing the AD in the zone where the
gravity of the MBH dominates. The velocity of a star, crossing the AD at a
distance $R$ is
\begin{equation}
\Vs=qc(\Rg/R)^{1/2}=qc(10^{-3}m_8/r_{-2})^{1/2},
\end{equation}
where $c$ is the velocity of light, $\Rg=2G\MBH/c^2$ is the gravitation
radius of MBH, $m_8=\MBH/10^8M_{\odot }$ is the MBH mass, $r_{-2}=R/(0.01~
{\rm pc})$ is the 
distance from the MBH in units of $0.01$ pc, and $1>q\geq 1/\sqrt{%
2}$ is the coefficient for the Keplerian orbits: $q$ is close to 1 for the
very stretched orbits, and $q=1/\sqrt{2}$ for the circular ones. In our
approximation of almost circular orbits we take $q=1/\sqrt{2}$, and $%
\Vs\simeq \sqrt{GM/R}$. So in convenient units 
$v_{\rm s9}=\Vs/(10^9~{\rm cm~s}^{-1})$,
\begin{equation}
v_{\rm s9}\simeq 0.67(m_8/r_{-2})^{1/2}.
\end{equation}

The normal and tangential velocity components with respect to the thin disk
are $V_{\rm n}=\Vs\sin i$ and $V_{\rm t}=\Vs\cos i$, where $i$ is the orbit-to-disk
inclination angle. The gas motion in the disk is also close to the circular
one, $V_{\rm d}\simeq \Vs$; so, the relative velocity of the star and gas in the
Keplerian disk is $V_{\rm re}=[V_{\rm n}^2+(V_{\rm d}-V_{\rm t})^2]^{1/2}$, 
and we obtain $%
v_{\rm re9}\simeq v_{\rm s9}2^{1/2}(1-\cos i)^{1/2}$. Note that this velocity is
supersonic in the disk with $\Td\sim 10^4$ K even if the inclination angle is
very small, down to $i\sim 0.1^{\circ}$.

It is obvious that the drag force acting at the star in the disk is
proportional to the cross-section of the star, the disk density, and the
square of the relative star-disk velocity, but the coefficient of the
proportionality is known with an accuracy of a factor of a few; hereafter we
will denote it with $Q$. Several attempts to estimate $Q$ (Zentsova 1983,
Ostriker 1983, $\dot Z$urek et al.1994) gave slightly different results. We
will use for the energy loss of the star crossing thin AD the expression
\begin{equation}
\Delta E=Q\pi \Rs^2\Sigd(\Vs)^2\phi (i),
\end{equation}
where $\Rs^{}$ is the radius of the star, $\Sigd$ is the surface mass
density of the AD, and $\phi (i)$  is the function of the inclination angle.
Within our approximations, one has $\phi (i)\sim
[(V_{\rm n})^2-(V_{\rm d}-V_{\rm t})^2]/(\Vs)^2=2\cos i
(1-\cos i)=2\sin i^2\cos i/(1+\cos i)$.

Physically this means that the star gives its velocity (in the system
co-rotating with disk) to the column of the disk matter with
cross-section close to that of the star, and with height equal to the disk
thickness. This brakes the normal component of the star velocity, but
accelerates the star in the direction of the disk rotation.

The energy generated by the drag force is
\begin{equation}
\label{eq:diss1}E_{\rm diss}\sim Q\pi \Rs^2\Sigd(V_{\rm re})^2=Q\pi \Rs^2\Sigma
_{\rm d}(\Vs)^22(1-\cos i)
\end{equation}
which differs from $\Delta E$ with $1/\cos i$; we will denote $2(1-\cos
i)=\phi_1(i)=\phi (i)/\cos i$.

Hereafter we will use the Solar units $\rs=\Rs/R_{\odot }$; $%
\ms=\Ms/M_{\odot }$; so, the cross section of a star is $\pi \Rs^2\simeq
1.54\cdot 10^{22}r_{\rm s}^2~{\rm cm}^2$. We also define the usual values $%
m_8=\MBH/10^8M_{\odot }$, $\dot m=(dM/dt)/\dot M_{\rm Ed}$, 
$\dot M_{\rm Ed}\simeq
m_8M_{\odot } {\rm yr}^{-1}$.

Supposing that the stellar orbits are close to circular, we can
approximate the time interval between two crossings of the disk by a star as
$\Delta T=\pi R/\Vs\simeq 1.4\cdot 10^8m_8^{-1/2}r_{-2}^{3/2} ~{\rm s}$.

Denoting $\Sigd=10^4\Sigma _4 ~{\rm g~cm^{-2}}$, one can define the energy
dissipated per crossing
\begin{equation}
\Delta E\simeq 0.7\cdot 10^{44}Q\rs^2\Sigma _4m_8r_{-2}^{-1}\phi_1(i) ~{\rm 
erg},
\end{equation}
and the averaged rate of the energy change of a star, $\Delta E/\Delta T$ is
\begin{equation}
\label{eq:sd1}dE/dt\simeq 0.5\cdot 10^{36}Q\rs^2\Sigma
_4m_8^{3/2}r_{-2}^{-5/2}\phi (i)~{\rm erg~s}^{-1}.
\end{equation}

The commonly accepted models for the AD are so-called $\alpha $-disks, the
Shakura-Sunyaev model (Shakura \& Sunyaev 1973, hereafter SS), and the
Novikov \& Thorne (1973) model (hereafter NT); the last was used in Rauch
(1995). On the other hand, physical assumptions of the models, the
geometrically thin and optically thick of the $\alpha $-disks theory, are
not universal. Other processes like energy advection, temperature
instability, self-gravity and the CSC gravity are essential in different
cases (see the recent review of Park \& Ostriker, 2001).

We will use mostly the NT model, as it contains a simple radial dependence
of the surface density of the AD; according to Rauch (1995) the surface density
of the AD is
\begin{equation}
\label{eq:NT}\Sigma _4\simeq 2.7\cdot 10^2m_8^{1/4}\alpha _{-2}^{-4/5}(%
\stackrel{\cdot }{m}_{-1})^{7/10}r_{-2}^{-3/4}.
\end{equation}

Then the energy dissipation per crossing is
\begin{equation}
\Delta E\simeq 1.9\cdot 10^{46}Q\rs^2m_8^{5/4}\alpha _{-2}^{-4/5}
{\dot m}_{-1}^{7/10}r_{-2}^{-7/4}\phi_1(i)~{\rm erg},
\end{equation}
and the average rate of the energy change of a star is
\begin{equation}
dE/dt\simeq 1.35\cdot 10^{38}Q\rs^2m_8^{7/4}r_{-2}^{-13/4}\phi (i) ~{\rm erg~s
}^{-1}.
\end{equation}

According to the SS model, the parameters of the accretion disk behave
differently in three regions (a, b, c), and we have from SS the positions
of a/b and b/c borders: $r_{\rm ab-2}=2.6m_8^{23/21}\alpha ^{2/21}(\dot
m\varphi _{\rm ab})^{16/21}$; 
$r_{\rm bc-2}=18.9m_8(\dot m\varphi _{\rm bc})^{2/3}$;
where $r_{\rm ab-2}=R_{\rm ab}/10^{-2}$ pc, $\varphi =1-r^{-1/2}$.

The surface density and the height scale in the three zones may be written
as
\begin{eqnarray}
a)\Sigma & \simeq & 2.8\cdot 10^4(\alpha\stackrel{\cdot }{m}\varphi
)^{-1}m_8^{-3/2}r_{-2}^{3/2} ~{\rm g~cm}^{-2}, \nonumber  \\
z &\simeq & 3.2\cdot 10^{14}\cdot {m}m_8\varphi ~{\rm cm},\\
b)\Sigma & \simeq & 2.1\cdot
10^5\alpha ^{-4/5}m_8^{4/5}(\stackrel{\cdot }{m}\varphi )^{3/5}r_{-2}^{-3/5},
\nonumber  \\
z & \simeq & 0.85\cdot 10^{14}\alpha ^{-1/10}m_8^{39/20}(\stackrel{\cdot }{m}%
)^{2/5}\varphi ^{1/5}r_{-2}^{21/20},\nonumber \\
c)\Sigma & \simeq & 3.11\cdot 10^5\alpha%
^{-4/5}(\stackrel{\cdot }{m}\varphi )^{7/10}m_8^{19/20}r_2^{-3/4}, \\
z & \simeq & 0.67\cdot 10^{14}\alpha ^{-1/5}m_8^{-1/10}(\stackrel{\cdot }{m}%
\varphi )^{3/20}r_{-2}^{9/8} \nonumber
\end{eqnarray}

With these relations, the energy dissipation per crossing in the a, b,and c
zones of the disk are (in $10^{45}$ erg):
\begin{eqnarray}
\lefteqn{a)\Delta E\simeq 0.2Q\rs^2(\alpha \stackrel{\cdot }{m})^{-1}m_8^{-1/2}r_{-2}^{1/2}\phi_1(i), \nonumber}
 \\
\lefteqn{b)\Delta E\simeq 1.47Q\rs^2\alpha ^{-4/5}(\stackrel{\cdot }{m})^{3/5}m_8^{9/5}r_{-2}^{-8/5}\phi_1(i)},
 \\
\lefteqn{c)\Delta E\simeq 2.18Q\rs^2\alpha ^{-4/5}(\stackrel{\cdot }{m})^{7/10}m_8^{39/20}r_{-2}^{-7/4}\phi_1(i). \nonumber}
\end{eqnarray}

The average energy losses of a star for the three zones are (in $10^{37}$erg s$^{-1}$):
\begin{eqnarray}
a)dE/dt & = & 0.14Q\rs^2(\alpha \stackrel{\cdot }{m})^{-1}r_{-2}^{-1}\phi (i) \nonumber \\
b)dE/dt & = & 1.1Q\rs^2\alpha ^{-4/5}(\stackrel{\cdot }{m}%
)^{3/5}m_8^{23/10}r_{-2}^{-21/10}\phi (i) \\
c)dE/dt & = & 1.6Q\rs^2\alpha ^{-4/5}(\stackrel{\cdot }{m}%
)^{7/10}m_8^{49/20}r_{-2}^{-3/4}\phi (i) \nonumber .
\end{eqnarray}

\subsection{The star-star interactions and the inflow of stars}

In spherically symmetrical conditions, the influence of the central
massive black hole on the stellar cluster structure and the inflow of
stars into the MBH was considered by Peebles (1972) and solved by Bahcall \&
Wolf (1976) (hereafter BW). BW showed that the presence of a MBH in a
dense stellar cluster typically leads to the creation of a central stellar
cusp with a density profile $n(R)=n_0(R/R_{\rm a})^q$, with $q\simeq -7/4$ and
$R_{\rm a}\simeq R_{\rm ss}\MBH/M_{\rm SS}$. 
The stellar distribution function depends
on the stellar energy as $f(E)\sim E^{1/4}$, where $E=G\MBH/r-1/2\Vs^2$
and $\Vs$ is the stellar velocity. Both the creation of the cusp and the inflow
of the stars into the MBH are due to star-star gravitational
interaction. The result was obtained for an isotropic stellar distribution
function, but BW showed that it is valid if we use anisotropic
distribution functions as well. However, in the case of AGNs with accretion
disks, the interactions of the stars with the disk in the central part of
the stellar system are inevitable, and the influence of the interaction on 
the stellar dynamics have to be taken into account.

Let us compare the rate of the energy changes of a typical star (let it be
the solar-type star) for the cases of star-star (hereafter s-s) 
interaction and star-disk interaction. For the former case one has
\begin{equation}
\label{eq:ss1}d\Ess/dt\simeq 4\pi G^2\Ms^3\Vsr^{-1}\ns \ln \Lambda ,
\end{equation}
where $\Vsr$ is the effective relative velocity of the stars, $\ns$ is
the stellar density and $\ln \Lambda $ is the Coulomb logarithm; for the
latter case Eq.~(\ref{eq:sd1}) applies.

It is essential to note that in the present work we will use some
effective values of physical variables. In principle these variables
show a certain distribution at each radius: stars at each radius have 
various inclinations, velocities relative to the disk etc. A complex
numerical approach would be necessary in order to model the distributions
of these quantities. Knowing the distributions we would be able to compute
accurately the mean or median values of the parameters. However, such a
program is far beyond the scope of the present paper. Instead, we  
reduce the dimension of the problem by
introducing a priori some effective values at each radius, 
with values following from processes we
consider. It permits us to determine
physically reasonable radial trends; but it is obvious that the
obtained results are true only in a qualitative sense.

If the mass of the CSC is larger than the mass of MBH, the effective
velocity of the stars in the outer parts of the CSC is close to the virial
velocity $\Vs \approx (GM_{\rm sc}/R_{\rm sc})^{1/2}$, where $M_{\rm sc}$ 
and $R_{\rm sc}$
are the mass and radius of the CSC. In the inner part of the AD the velocity
at the almost circular orbits is close to $\Vs\approx (GM/R)^{1/2}$, where
$M(R)=\MBH+M_{\rm sc}(R)+M_{\rm AD}(R)$ is the total mass within the radius $R$.

Rauch (1995) has shown that close to a black hole the characteristic
two-body relaxation time $t_{\rm ss}\sim v^3/(\pi G^2 \Ms^2 \ns)\sim
10^{12}(\ns/10^7 ~{\rm pc}^{-3})$ is much longer than the characteristic time
of the stellar orbit alignment with the disk, $t_{\rm i}\sim t_{\rm orb}\Ms/(\pi
\Rs^2)/\Sigd$. This is still true in our case of more dense CSC if it is
spherically symmetric, which can be seen from the numerical estimates of
the correspondent energy-loss rates (in erg~s$^{-1}$):
\begin{eqnarray}
d\Ess/dt & \simeq & 5.9\cdot
10^{32}\ms^3m_8^{-1/2}r_{-2}^{1/2}n_{10}(\ln \Lambda )_1, \nonumber \\
d\Esd/dt & \simeq & 5\cdot 10^{35}Q\Sigma _4\rs^2m_8^{3/2}r_{-2}^{-5/2},
\end{eqnarray}
where $m_8=M(R)/10^8~M_{\odot }$, $n_{10}=\ns/(10^{10}~{\rm pc}^{-3})$.

One can see that the star-disk interaction is relatively strong even in the
outer parts of the AD. We can estimate the critical outer radius of the
region where the two interactions became comparable, if the inclination of
the orbits is negligible [$\sin(\ief)\sim 1$] in this outer region:
\begin{equation}
\label{eq:rcrit}r_{-2}^{\rm crit}\sim 9.5\Sigma
_4^{1/3}\rs^{2/3}\ms^{-1}m_8^{2/3}[n_{10}(\ln \Lambda )_1]^{-1/3},
\end{equation}
that is $R^{\rm crit}\simeq 0.1$ pc (if $\Sigma _4^{}=m_8=n_{10}=1$).

At smaller radii the inelastic star-disk collisions become important, and
the stellar orbits incline to the disk. If the orbit evolution starts
from a non-inclined orbit, the orbit will quickly decline to the disk plane
(as was shown by Rauch 1995). But when the inclination angle becomes
small, the situation can change due to the increased stellar density. 
The reason is that the inclination of the orbits influences the
relative velocities, and consequently the relaxation times of the star-disk
and star-star collisions.

We will assume that the inclination angles of the orbits are smoothly
distributed above the disk, and the distribution can be characterized by
some effective inclination angle $\ief$. The characteristic relative
velocities of stars in the co-rotating (with the disk) coordinate system are
then close to $\Vsr\simeq \Vs\sin \ief$, that is close to the average
velocity component normal to the disk. Introducing the
effective inclination angle into the expressions of the energy changes due to
the star-star and star-disk interactions, we have from Eqs.~(\ref{eq:sd1})
and (\ref{eq:ss1}) numerically (in erg~s$^{-1}$):
\begin{equation}
\label{eq:ss2}d\Ess/dt\simeq 5.9\cdot
10^{32}\ms^3m_8^{-1/2}r_{-2}^{1/2}n_{10}(\ln \Lambda )_1/\sin(\ief),
\end{equation}
\begin{equation}
\label{eq:sd2}d\Esd/dt\simeq 5\cdot 10^{35}Q\Sigma
_4\rs^2m_8^{3/2}r_{-2}^{-5/2}\phi (\ief).
\end{equation}

The important point is that the s-s interaction behaves as $1/\sin(\ief)$ and
the s-d interaction is proportional to $[\sin(\ief)]^2$, so both
interactions may be comparable at small angles. It is intuitively clear,
that an equilibrium situation exists when $d\Ess/dt\simeq d\Esd/dt$%
. The process is similar to the Brownian motion of the normal velocity
component $V\sin i$, and the equilibrium velocity dispersion corresponds to
an equilibrium of braking and fluctuation forces in the Langevin
equation.

As the characteristic time of approaching the equilibrium is relatively
short (close to the characteristic time of orbit inclination, $t_{\rm i}\simeq
t_{\rm dyn}(\Ms/\pi \Rs^2)/\Sigd$), we will suppose further that the
equilibrium condition $d\Ess/dt\simeq d\Esd/dt$ is realized at any
distance $R< R^{\rm crit}$. This equilibrium means that in a dense enough CSC
the stars will never be completely incorporated into the disk plane: when
orbits are inclined too much, the pair stellar interactions will come into
play, and they will scatter the stars to less inclined orbits (the
eccentricities of the orbits will also be scattered to the same order of
magnitude $e_{\rm ef}\sim \sin i$). The immediate consequence of this property,
which leads to the fluctuation of the inclination angle around some
effective value, is that the fluctuating drag force maintains some constant
dissipative flow of the stars to the center of the disk. The rate of the
inflow is
\begin{equation}
\VR=dR/dt=dR/dE(dE/dt),
\end{equation}
where $dE/dR=G\MBH\Ms/R^2 ~{\rm erg~cm}^{-1}$, that is $dR/dE\simeq
3.37\cdot10^{-35}\ms^{-1}m_8^{-1}r_{-2}^2$ cm~erg$^{-1}$, and $dE/dt$ is the
averaged dissipative energy loss of the star. We therefore obtain
\begin{equation}
\VR\simeq 17Q\Sigma _4\rs^2\ms^{-1}m_8^{1/2}r_{-2}^{-1/2}\phi (i)~{\rm cm~s}
^{-1}.
\end{equation}

In stationary conditions, the flow of the stars is
\begin{equation}
d\ns/dt=\dot n\simeq 4\pi \ns\VR R^2 \sin(\ief),
\end{equation}
where $\ns$ is the stellar density at the AD plane and $\ief(R)$ is the
effective inclination angle of stellar orbits, which means also the part
of the solid angle where the inflow is concentrated.

Numerically, this formula gives the flow:
\begin{equation}
\label{eq:ndot1}\ndoty\simeq 2\cdot 10^{-3}n_{10}Q\Sigma
_4\rs^2\ms^{-1}m_8^{1/2}r_{-2}^{3/2}\sin \ief\phi (\ief)~{\rm yr}^{-1}
\end{equation}
in units of stars/year, where $n_{10}$ is the stellar density in
units $10^{10}~{\rm pc}^{-3}$.

\section{The distribution of stars in the central part of CSC}

\subsection{Constant stellar masses and inflow of the stars}

As was shown in the previous section, the stellar orbits tend to decline
to the disk and to shrink due to the energy loss of the stars crossing the
disk, but the inclination of the orbits is limited by the scattering due to
the gravitational star-star interaction. In the stationary case both
processes are in equilibrium, which defines the inflow of the stars to
the center.

This process can be described with a system of two equations, which permit
us to find the distribution of the stars. The first equation is the star
inflow Eq.~(\ref{eq:ndot1}), and the second is the condition of
equilibrium of the inelastic (star-disk) and elastic (star-star) collisions,
\begin{equation}
\label{eq:sdss1}d\Esd/dt=d\Ess/dt.
\end{equation}

Eq.~(\ref{eq:ndot1}) can be written in the form
\begin{equation}
\label{eq:sin1}\sin (\ief)\phi (\ief)n_{10}\simeq 5\cdot 10^2\dot
n_{\rm y}/(Q\Sigma _4)\ms\rs^{-2}m_8^{-1/2}r_{-2}^{-3/2}
\end{equation}
where $\ndoty$ is the stellar inflow in the units 1/year, $n_{10}$ is the
stellar density in $10^{10}~{\rm pc}^{-3}$. 
On the other hand, from 
Eq.~(\ref{eq:sdss1}), using Eqs.~(\ref{eq:ss2}) and (\ref{eq:sd2}) one obtains:
\begin{eqnarray}
\lefteqn{\sin (\ief)\phi (\ief)/n_{10}\simeq} \nonumber\\ & 1.2\cdot
10^{-3}(Q\Sigma  _4)^{-1}\ms^3\rs^{-2}m_8^{-2}r_{-2}^3(\ln \Lambda )_1.
\label{eq:sin2}
\end{eqnarray}

Eqs.~(\ref{eq:sin1}) and (\ref{eq:sin2}) form a system of
equations for two variables: $n_{10}(r)$ and $\ief(r)$. Dividing Eq.~(\ref
{eq:sin1}) by (\ref{eq:sin2}) one has
\begin{equation}
\label{eq:n10}n_{10}(r)\simeq 0.6\cdot 10^3[{\ndoty}/(\ln \Lambda
)_1]^{1/2}\ms{}^{-1}m_8^{3/4}r_{-2}^{-9/4}.
\end{equation}

This distribution is independent of the disk parameters, paradoxically at
first glance. The reason is the underlying assumption of the equilibrium
of non-elastic (s-d) and elastic (s-s) processes, but this condition is
satisfied only in dense enough disks, so it is not valid for low surface
density disks.

Multiplying Eq.~(\ref{eq:sin1}) by (\ref{eq:sin2}), and using (for $\ief<\pi /3$ )
the approximation $\phi (i)=2\sin ^2i\cos i/(1+\cos i)\approx \sin ^2i$, one
has
\begin{eqnarray}
\lefteqn{\sin (\ief)\approx}\nonumber\\ & &0.9[\ndoty(\ln\Lambda
)_1]^{1/6}(Q\Sigma  _4)^{-1/3}(\ms/\rs)^{2/3}m_8^{-5/12}r_{-2}^{1/4}
\label{eq:ieff}\end{eqnarray}

This solution is valid in the innermost parts of the CSC/AD system, where $%
\sin (\ief)<1;$ in the outermost parts the orbits are non-circular and we
pose $\sin (\ief)=1$ there.

If we use the disk surface density $\Sigma $ of the NT disk
from Rauch (1995), then
\begin{eqnarray}
\lefteqn{\sin (\ief)  \approx  0.14[\ndoty(\ln\Lambda
)_1]^{1/6}(Q)^{-1/3} \times} \nonumber \\
 & & (\ms/\rs)^{2/3}m_8^{-1/2}\alpha
_{-2}^{4/15}(\stackrel{ \cdot }{m}_{-1})^{-7/30}r_{-2}^{1/2}.
\end{eqnarray}

The dependences of $\sin \ief$ on $r$ for three zones of the SS accretion
disk are:
\begin{eqnarray}
\lefteqn{  \sin (\ief)(r)\simeq} \nonumber \\
\lefteqn{ a)0.64Q^{-1/3}(\ms/\rs)^{2/3}(\alpha \stackrel{\cdot }{m})^{1/3}m_8^{1/12}[{\dot n}_{\rm y}(\ln\Lambda )_1]^{1/6}r_{-2}^{-1/4}} \nonumber \\
\lefteqn{ b) 0.34Q^{-1/3}(\ms/\rs)^{2/3}(\alpha )^{4/15}(\stackrel{\cdot }{m})^{-1/5}m_8^{-41/60} \times } \nonumber \\ & & [{\dot n}_{\rm y}(\ln\Lambda )_1]^{1/6}r_{-2}^{9/20} \nonumber \\
\lefteqn{ c)
0.3Q^{-1/3}(\ms/\rs)^{2/3}(\alpha )^{4/15}(\stackrel{\cdot }{m})^{-7/30}m_8^{-11/15} \times} \nonumber \\  &  & [{\dot n}_{\rm y}(\ln\Lambda )_1]^{1/6}r_{-2}^{1/2}.
\end{eqnarray}

The dependence of $\sin(\ief)$ and $n_{10}$ on $r_{-2}$ for different $
\Sigma _4(r)$ (the SS, NT and constant density disk cases) are shown in
Fig. 1. We adopt $\MBH=10^8M_{\odot }$, $\alpha =0.01$ and $\dot m=1$. One
can see that the stellar density rises quickly at small radii, while the
effective inclination angle of the stellar orbits diminishes to the center.
As can be seen from the picture, the difference in the behavior of $
\sin(\ief)$ for different disk models is not too large (though it has a
wide minimum in the case of an SS disk), so further on we will use the NT model
for our model calculations.

\begin{figure}
 \epsfxsize=0.45\textwidth \epsfbox{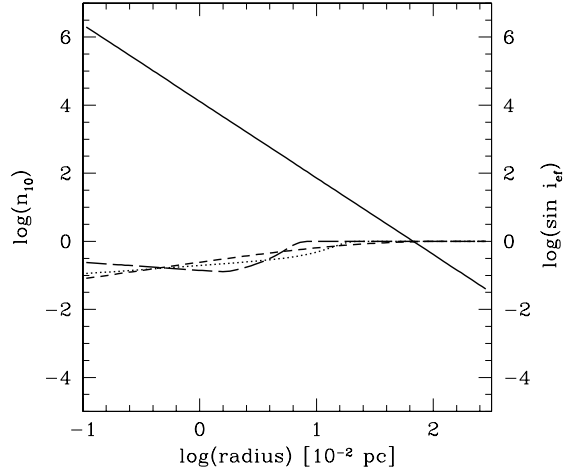}
 \caption {Radial dependence of stellar density $n_{10}$ (continuous line) and $\sin (\ief$
in the case of constant stellar flow for SS disk, constant surface density disk and NT disk (long-dashed, short dashed and dotted line, correspondingly}.
\end{figure}

The inflow rate of the stars ${\dot n}_{\rm y}$ in this consideration (in the
approximate model we investigate here) is assumed to be given by a constant, 
which is
determined by the conditions at the outer border, $R=R_{\rm out}.$ This outer
border is close to the above-estimated value $R^{\rm crit}$, determined by 
condition (\ref{eq:rcrit}). Physically, the value $\dot n(R_{\rm out})$ is
determined by the diffusion process, which depends on the properties of the
compact stellar cluster, and which for the case of a non-rotating CSC was described
by Bahcall \& Wolf (1976). It is clear that in the central part of the CSC
the stellar density can be so large that physical collisions of the
stars can occur; besides, the stars can lose mass when they cross the disk.
These processes have to be taken into account.

\subsection{The contact stellar collisions}

Let us consider the change of stellar flow due to the contact (sometimes
it is called `physical') collisions of the stars; this is the non-elastic
s-s process. The collision velocities are about $V_{\rm c}\sim V\sin i$ and the
escape velocity from the surface of a star can be estimated as $v_{*}\simeq
6\cdot 10^7[\ms/\rs]^{1/2}~{\rm cm~s}^{-1}$. If $V_{\rm c}>v_{*}$, the averaged
mass loss per collision is large (Murphy et al. 1991), and stars
are completely disrupted if the impact parameter is less than some value. In
the opposite case $V_{\rm c}<v_{*}$ (if, say, $\sin(\ief)<<1$), the collisions
can produce binary stars, or stars can be merged. Within the framework of
the present paper we adopt equal masses of stars (which itself is far enough
from reality!). In contact collisions a fraction of mass is always lost,
but merging can increase stellar mass. Therefore, we will assume that both
processes somehow compensate, and consider unchanged stellar masses, but
take into account a decrease of the stellar density $\ns$ and the stellar
inflow $\ndoty$ as a result of the direct s-s collisions.

In Sect.~3.1 we have obtained formulae for the stellar density $\ns$
and $\sin (\ief)$, depending on the inflow of stars $\ndoty$. These
formulae are valid also for the case $\ndoty^{}=\ndoty^{}(R)$. The
frequency of the physical collisions of a given star with another stars (all
stars have stellar radii $\Rs$) is
\begin{equation}
\nu^1_{\rm ss}=4\pi \ns\Rs^2\Vs\sin (\ief),
\end{equation}
and numerically:
\begin{eqnarray}
\lefteqn{\nu^1_{\rm ss}\simeq 0.4\cdot 10^{-12}[{\dot
n_{\rm y}}(\ln\Lambda  )_1]^{1/2} \times }\nonumber\\ & &
\ms^{-1}\rs^2m_8^{5/4}r_{-2}^{-11/4}\sin(\ief) ~{\rm s}^{-1}.
\label{eq:dnss}
\end{eqnarray}
The frequency of the star-star physical (contact) collisions in the
spherical layer is
\begin{equation}
d\nu _{\rm ss}=16\pi ^2\Rs^2\ns^2\Vsr^2[\sin (\ief)]^2dR,
\end{equation}
thus giving,
\begin{eqnarray}
\lefteqn{d\nu _{\rm ss}\simeq 0.95\cdot 10^{-4} \times }\nonumber\\
& &\rs^2\ms^{-2}m_8^2\dot
n_{\rm y}(\ln\Lambda  )_1^{-1}r_{-2}^{-3}\sin^2(\ief)dr_{-2}~{\rm s}^{-1}.
\end{eqnarray}

The rate of the diminishing of the stellar inflow due to disruption of stars
is $d({\ndoty})/dr=-q_{\rm v}d\nu_{ss}(1/{\rm s})(3\cdot 10^7 {\rm s}/{\rm yr})$, 
where $q_{\rm v}$ is
equal to $2$ or $1$, for the disruptive or merging case, depending on the 
$V_{\rm r}/v_{*} $ 
relation. Besides, the cross-section of the stellar collisions
depends on the relative stellar velocity $V_{\rm r}/v_{*}$ due to gravitational
focusing at small velocities. We use these dependences in the numerical
calculations (see below). Here we will neglect them and obtain approximate
solutions with $q_{\rm v}=1.$

In the spherical layer between $R$ and $R+dR$
\begin{eqnarray}
\lefteqn{
d(\ndoty)\simeq } \nonumber \\
& 2.9\cdot 10^3{\ndoty}(\ln\Lambda
)_1^{-1}\ms{}^{-2}\rs^2m_8^2\sin ^2(\ief)r_{-2}^{-3}dr_{-2}.
\end{eqnarray}
Using $\sin (\ief)$ from Eq.~(\ref{eq:ieff}), one has ($\ndoty$ in units yr$^{-1}$)
\begin{eqnarray}
\lefteqn{
d(\ndoty)\simeq 1.25\cdot 10^3{\ndoty}^{4/3} \times }\nonumber\\ & &(Q\Sigma
_4)^{-2/3}(\ln\Lambda
)_1^{-2/3}\ms{}^{-2/3}\rs^{2/3}m_8^{7/6}r_{-2}^{-5/2}dr_{-2}.
\end{eqnarray}

The solution of the differential equation can be obtained for different $%
\Sigma (r)=Ar^u$. If we assume $\Sigma =const$, the solution is
\begin{eqnarray}
\lefteqn{
\ndoty(R)\simeq \{\dot n_{\rm 0y}^{-1/3}+0.26\cdot 10^3(Q\Sigma
_4)^{-2/3} \times }\nonumber\\ & & (\ln\Lambda
)_1^{-2/3}\ms{}^{-2/3}\rs^{2/3}m_8^{7/6}[r_{-2}^{-3/2}-r_{0-2}^{-3/2}]
\}^{-3}.
\end{eqnarray}

If we use $\Sigma $ from the NT model, then
\begin{eqnarray}
\lefteqn{ d[\ndoty(R)]\simeq 0.5\cdot 10^2[\ndoty(R)]^{4/3}(Q)^{-2/3}
(\ln\Lambda  )_1^{-2/3} \times } \nonumber \\
 & &  \ms^{-2/3}\rs^{2/3}m_8^{5/6}\alpha _{-2}^{8/15}(\dot m_{-1})^{-7/15}r_{-2}^{-2}dr_{-2},
\end{eqnarray}
and the approximate solution is
\begin{eqnarray}
\lefteqn{
\ndoty(R)\simeq \{\dot n_{\rm 0y}^{-1/3}+3.4(Q)^{-2/3}(\ln \Lambda
)_1^{-2/3}\times}\nonumber\\ & &\ms{}^{-2/3}\rs^{2/3}m_8^{5/6}\alpha _{-2}^{8/15}
(\dot m)^{-7/15}[r_{-2}^{-1}-r_{0-2}^{-1}]\}^{-3}.
\end{eqnarray}

With these results and using Eqs.~(\ref{eq:n10}) and (\ref{eq:ieff}), we have the
stellar density and the effective inclination angle of orbits
\begin{eqnarray}
& n_{10}(R)\approx
  0.67\cdot 10^3[{\ndoty(R)}/(\ln \Lambda%
)_1]^{1/2}m_8^{3/4}r_{-2}^{-9/4}m_{\rm s}^{-1}, \nonumber \\
\lefteqn{
\sin (\ief)(R)\approx 0.93[\ndoty(R)(\ln \Lambda )_1]^{1/6}\times } \nonumber \\
 & (Q\Sigma %
_4)^{-1/3}(\ms/\rs)^{2/3}m_8^{-5/12}r_{-2}^{1/4}.
 \end{eqnarray}

A more precise result can be obtained numerically if we take into account the
dependence of the cross-section of the stellar collisions on the relative
stellar velocity due to gravitational focusing at small velocities $
V_{\rm rel}\leq V_{\rm esc}\simeq 6\cdot 10^7(\ms/\rs)^{1/2}$ cm s$^{-1}$. 
These cases take
place at small $\sin \ief$ and/or at large distances, where we have to use
collision cross sections as 
$\sigma _{\rm ss}=4\pi \Rs^2(1+$ $V_{\rm esc}/V_{\rm rel})^2$.

\begin{figure}
 \epsfxsize=0.45\textwidth \epsfbox{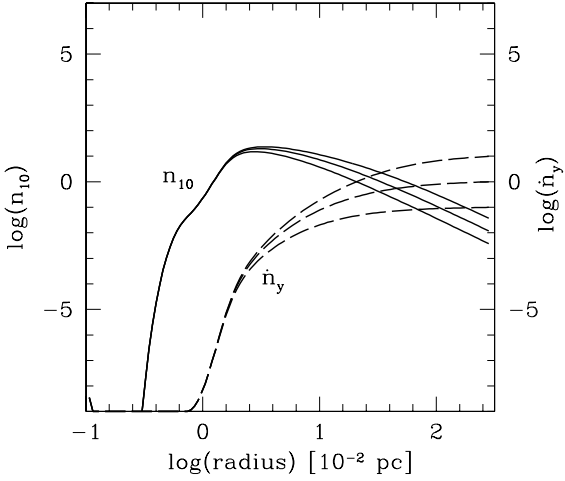}
 \caption{The radial dependence of the stellar density $n_{10}$,
and stellar flow $\ndoty$ in the case of  $\MBH=10^8~M_{\odot}$ and
the values of the initial stellar flow equal 0.1, 1.0 and 10 stars/yr 
(dashed lines).}
  \end{figure}

\subsection{Stellar mass loss due to s-d interactions}

In the above considerations we neglected the change of the masses of stars
due to s-s collisions, assuming that mass loss and stellar merging
compensate and that stellar masses keep their initial values. This approximation is
reasonable at large radii since the stellar density is small, and
collisions of stars are rare. On the other hand, stars would systematically
lose mass when crossing the disk, and this process is more intense at
small radii.

The frequency of the star-disk interactions of a single star in a circular
orbit with radius $R$ is inversely proportional to the half-period, $\nu
_{\rm sd}=1/\Delta T(R)$; numerically
\begin{equation}
\nu^1_{\rm sd}\simeq 0.71\cdot 10^{-8}m_8^{1/2}r_{-2}^{-3/2} {\rm s}^{-1}.
\end{equation}

The differential frequency of the star-disk crosses by stars from the volume
of a spherical layer $d{\it V}=4\pi R^2\sin (\ief)dR$ is
\begin{equation}
d\nu _{\rm sd}=4\pi \ns\Delta T^{-1}R^2\sin (\ief)dR,
\end{equation}
and expressed in [s$^{-1}$] gives
\begin{equation}
\label{eq:nusd}d\nu _{\rm sd}\simeq 0.6\ms^{-1}m_8^{5/4}[\ndoty/(\ln \Lambda
)_1]^{1/2}r_{-2}^{-7/4}\sin (\ief)dr_{-2}.
\end{equation}
As the effective inclination angle is non-zero (and typically it is larger
then several degrees), the star-disk velocities are supersonic and the stars
lose their masses when crossing the disk. Let us suppose that the mass-loss
rate is proportional to the dissipated energy
\begin{equation}
d\Ms/dt=q_{\rm m}(dE1/dt)(dM/dE),
\end{equation}
where $dE1/dt$ is the rate of the energy dissipation in the disk and 
$d\Ms/dE=-\Rs/(G\Ms)$.

Obviously, the envelopes of the red giants will be blown off in the process
of crossing the outermost parts of a typical accretion disk of the
bright AGN, so further on we will consider only the main-sequence stars.
Denoting $q_{\rm m}\sim 10^{-2}\qm2$, and using Eqs.~(\ref{eq:diss1}) and
(\ref{eq:sd1}) we have (in units of ${\rm g~s}^{-1}$)
\begin{equation}
d\Ms/dt=-2.6\cdot 10^{18}\qm2\rs^3\ms^{-1}Q\Sigma
_4m_8^{3/2}r_{-2}^{-5/2}\phi_1(\ief).
\end{equation}
We can deduce the radial dependence of $\ms$ from 
$d\Ms/dR={(d\Ms/dt)}/{V_{\rm r}}$. 
Assuming $\cos \ief\simeq 1$) and $\rs\simeq \ms$ (appropriate for the
main-sequence stars), one has
\begin{equation}
d\ms/\ms\simeq -2.3\qm2 m_8(dr_{-2}/r_{-2}^2).
\end{equation}
The solution of the equation is
\begin{equation}
\ms\simeq m_{\rm s0}\exp \{-K[1/r_{-2}-1/r_{-2}^{(0)}]\}.
\end{equation}
where $K=2.3\qm2 m_8$, and $r_{-2}^{(0)}=r_{-2}^{\rm crit}$ is the outer
(critical) radius of the stellar inflow. The peculiarity of the
process is that (within our approximations!) it does not depend on the 
$\Sigma _4(r)$ behavior.

In the region $r_{-2}<<r_{-2}^{(0)}$ the solution is $\ms\simeq m_{\rm s0}\exp
\{-K/r_{-2}\}$, and one can see that the mass of a star quickly goes to zero
when $r_{-2}<K$. Physically this means that at this distance the time scale
of the mass loss becomes less than that of the radial drift (inflow), so the
stars with diminished masses crowd close to $r_{-2}\sim K$. The
distance $R_{\rm m}\sim 0.01 K\qm2~{\rm pc}$ is the radius of melting of
the stars due to the star-disk crossings.

It can be noted that the time between consecutive crossings is several years
at this distance ($m_8=1$), and the duration of crossing is about one day,
so the stars really can be bloated due to deformation and heating of the
surface, and this leads to even quicker melting of the stars.

\begin{figure*}
 \parbox{\textwidth}{
   \parbox{0.3\textwidth}{\epsfxsize=0.3\textwidth\epsfbox{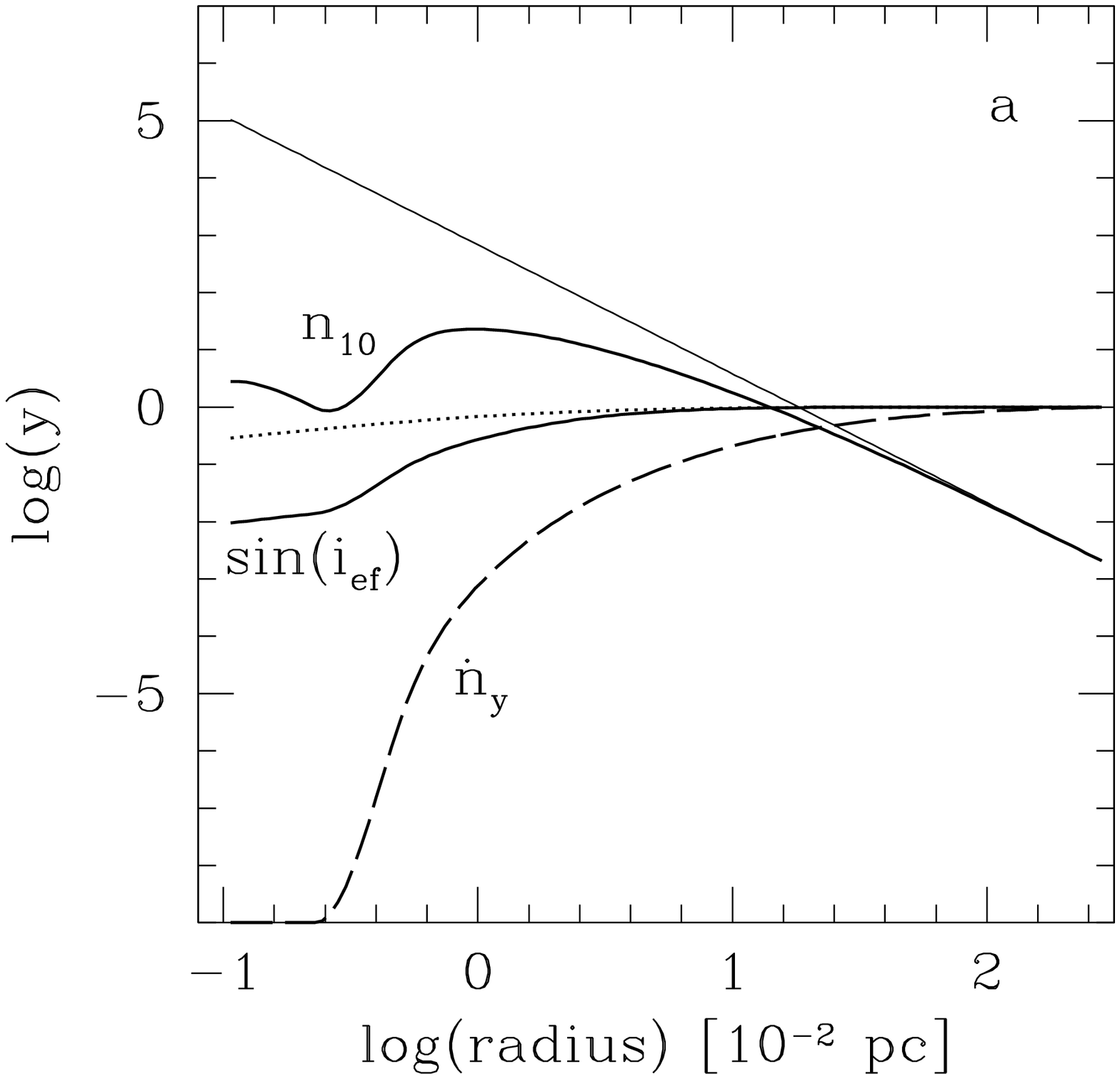}}
   \parbox{0.3\textwidth}{\epsfxsize=0.3\textwidth\epsfbox{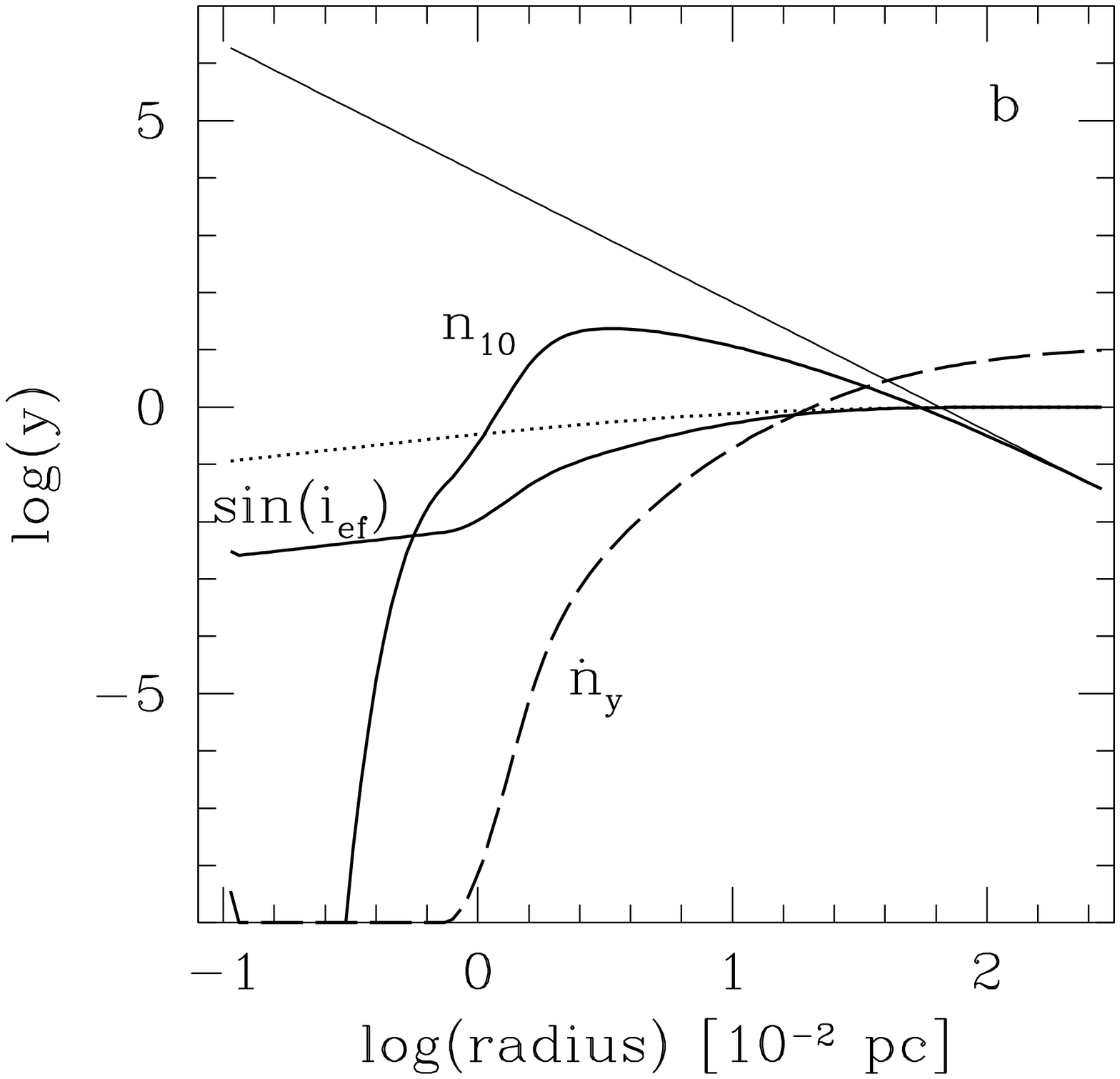}}
   \parbox{0.3\textwidth}{\epsfxsize=0.3\textwidth\epsfbox{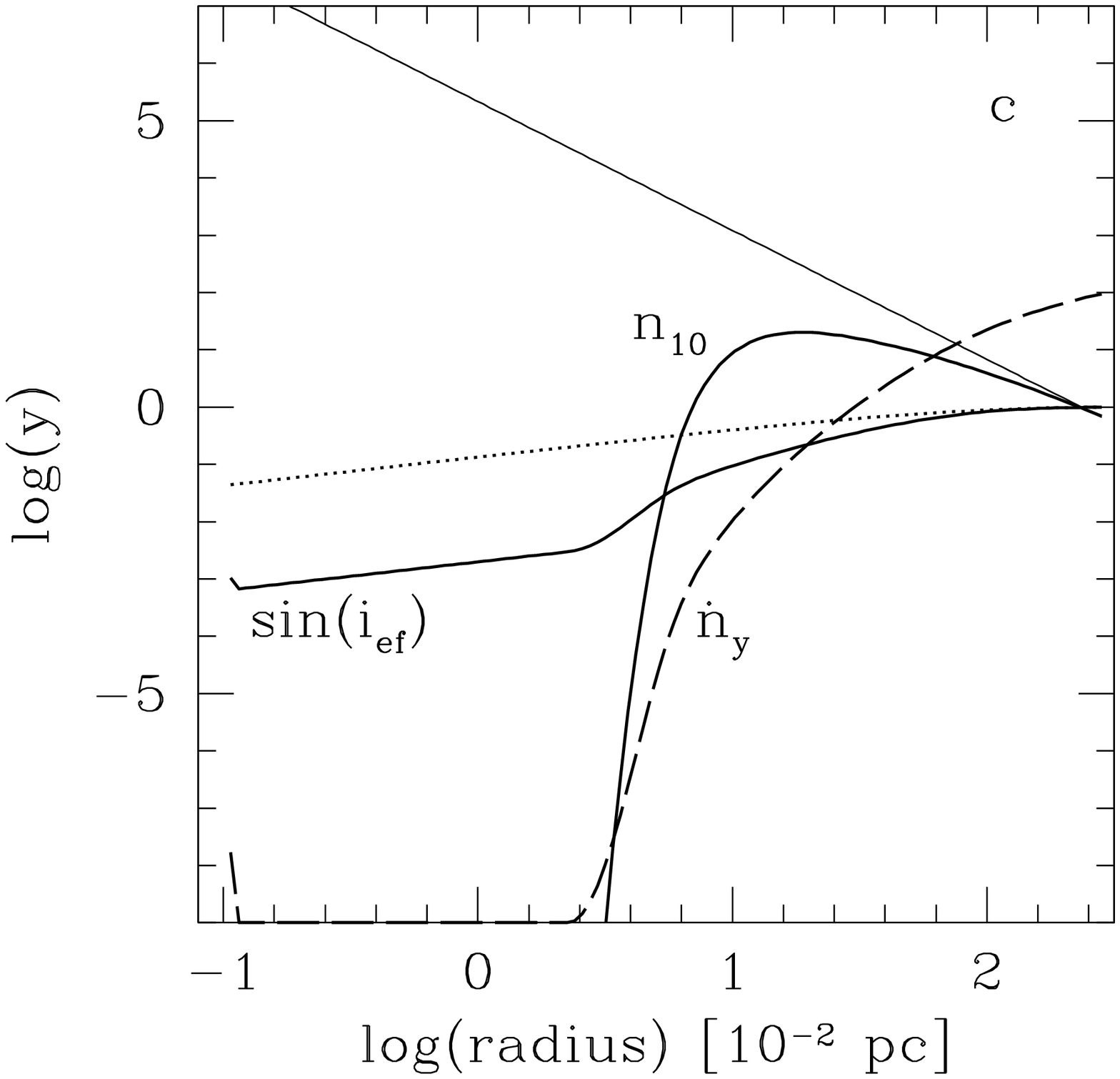}}
}
\caption{
The radial dependence of the stellar density $n_{10}$,
$\sin (\ief)$ and stellar flow $\ndoty$ for different black hole masses:
a)$\MBH=10^7M_{\odot}$, b)$\MBH=10^8M_{\odot}$, and
c)$\MBH=10^9M_{\odot}$. Thin continuous line shows the $n_{10}$ distribution in the absence of stellar collision effect.}
\end{figure*}

Adopting a minimal stellar mass of $\sim 0.01M_{\odot }$, one has $%
(r_{-2})_{\min }\sim -K/\ln (m_{\rm s0}/\ms),$ that is $(r_{-2})_{\min }\sim
0.5q_{\rm m-2}m_8$ for $m_{\rm s0}=1$, and $(r_{-2})_{\min }\sim \qm2m_8$ for $
m_{\rm s0}=0.1$ .

As the stellar density drops quickly due to the melting of stars at 
$r_{-2}<(r_{-2})_{\min }$, in the numerical calculation we used the
approximation
\begin{eqnarray}
\lefteqn{
n_{10}\simeq 1.3\cdot 10^3[{\ndoty}/(\ln\Lambda )_1]^{1/2}\times}\nonumber\\ & &m_8^{3/4}r_{-2}^{-9/4}m_{\rm s0}^{-1}\exp
\{[K/r_{-2}(1-0.5K/r_{-2}]\},
\end{eqnarray}
that is the stellar density is enhanced close to $r_{-2}\approx
2.3q_{\rm m-2}m_8 $ and then drops at smaller radii. This means the appearance of
some `stellar hoop' with the radius $r_{-2}=2.3\qm2m_8$. The numerical
calculations of the emission line profiles (see Sect.~\ref{sect:BELR})
give reasonable results with $\qm2\simeq 1$, which we accept below as the
standard value.

The influence of the process of star melting at small radii was taken
into account in the numerical calculations of the models, and the results
are presented in Figs. 2 and 3.

Our approximation works for sufficiently dense disks, so we show the results of
the numerical calculations for the case of NT disk with $\alpha =0.01$ and $%
\dot m=1$.

In Fig. 2 we show the radial dependences of stellar densities and stellar
flows for the case of $\MBH=10^8M_{\odot }$ with different initial (at the
outer border) values of the stellar flow 
$\dot n_{\rm y0}=0.1, 1, 10 ~{\rm yr}^{-1}$. One can see that at 
small distances all the flows $\ndoty(R)$ become
almost the same and independent on the initial values of $\dot n_{\rm y0}$ at
large radius. This result is notable, as it shows that in fact the
accretion disk itself can be built with the gas resulting from the s-s
disruptive collisions in the dense enough CSC.

We used this hypothesis for calculation of the reference models of
different AGNs, assuming that the accretion rate onto the MBH is always equal
to the initial mass inflow by stars at the outer border of the CSC/AD
system, $\ndoty^{}\ms=m_8\dot m$. In Fig. 3 we show the radial dependences
of stellar densities, inclination angles and stellar flows for the
reference models with $\MBH=10^7M_{\odot },10^8M_{\odot },10^9M_{\odot
}$; also cases without contact stellar collisions and mass loss due to
the disk crossing are shown for comparison (thin lines), and the behavior of
the stellar inflow $\ndoty(R)$ is marked (dashed lines).

\section{Broad emission line region}

\label{sect:BELR}

There is still no commonly accepted physical idea about the nature of the
broad emission line regions (BELR) of AGNs. The possible role of the
accretion disk was stressed by Dumont \& Collin-Souffrin (1990) and it was
subsequently considered in a number of papers (e.g. Collin \& Hure 2001).
There were, however, also many attempts to involve stars (viz.
excited, bloated or comet-like stars) in the explanation of
problems of clouds, moving and emitting in BELRs (Edvards 1980; Shull 1983;
Penston 1988; Scoville \& Norman 1988, 1995; Kazans 1989; $\dot Z$urek et
al. 1994). The most recent was the study by Alexander \& Netzer (1994, 1997),
where the authors considered emission from the outer envelopes of excited
stars and used a spherically symmetric model of the structure and dynamics
of the CSC in the vicinity of the MBH (Murphy et al. 1991). The application
of a spherical model to the stellar dynamics in BELRs is questionable,
because it completely ignores the interaction of stars with the AD. Our
model of stellar dynamics includes both interactions which together
determine the correct spatial and kinematical distribution of stars and we
can now built a model of the BELR based on that knowledge.

To define the physical nature of the BELR clouds, and the model of the
star-disk interaction in more details, we refer to the paper by 
$\dot Z$urek et al. (1994), where the structure of the plasma
around the star moving through the disk is presented as a combination of a
hot bow-shock filled with hot plasma, and a more dense and cold
wake, or tail, behind the star. We suppose that when the star comes
out from the disk surface, the appearance of the bow-shock from the surface
of the disk can be treated as a fireball, and the stellar wakes can be
identified with the clouds of the BELR.

$\dot Z$urek et al. (1994) assumed that the BELR clouds
are actually wakes of matter, dragged by the stars from the
disk. We accept this idea in a slightly changed variant, taking the view that
the matter dragged from the disk creates a fireball, and the wakes
contain the mass, lost by the stars when they cross the disk (both parts are
the same to within an order of magnitude). As was described in the previous
section, the fraction of mass lost by the star is proportional to the dissipated
energy $\Delta E$, that is $\Delta M_{\rm s}=0,01\qm2\Delta EdM/dE$, where $%
d\Ms/dE=\Rs/(G\Ms)$.

Taking $\Delta E$ from Eq.~(\ref{eq:sd1}), one has 
\begin{equation}
M_{\rm t}\simeq \Delta \Ms\simeq $
$3.5\cdot 10^{26}\qm2Q\Sigma _4\rs^3\ms^{-1}m_8r_{-2}^{-1}\phi_1 ~{\rm g},
\end{equation}
where $\qm2=1$ means that we assume that $1\%$ of the star's energy
dissipated in the disk is used for the mass loss of the star.

As the balance of forces acting on the wakes depends on the parameters
of AGNs and is model-dependent, we assume here the simplest case: the wakes
follow the stars. Let the wakes be cylinders with lengths approximately
equal to the height of the stellar orbit above the disk, $l_{\rm t}\sim R\sin
(\ief)$, and let the radii of the cylinders be close to the stellar
radius, $r_{\rm t}\sim \rs R_{\odot }$ (this last suggestion may require the
presence of a hot corona to provide the cold gas confinement). Then the mean
particle density in the cylinder is
\begin{equation}
n\simeq M_{\rm t}/[l_{\rm t}\pi \rs^2\sin (\ief)m_{\rm p}],
\end{equation}
or numerically
\begin{equation}
n\simeq 3.8\cdot 10^{11}q_{-2}Q\Sigma _4\rs\ms^{-1}m_8r_{-2}^{-2}\phi_1
/(\sin \ief)~{\rm cm}^{-3},
\end{equation}
and the column density is
\begin{equation}
N\simeq 6.5\cdot 10^{22}q_{-2}Q\Sigma _4\rs\ms^{-1}m_8r_{-2}^{-2}\phi_1
/(\sin \ief)~{\rm cm}^{-2}.
\end{equation}

The covering factor (CF) of a single cloud (along the disk radius) is
\begin{equation}
\Omega =2\Rs l_{\rm t}/4\pi R^2\simeq 3.7\cdot 10^{-7}\rs\sin \ief/r_{-2},
\end{equation}
and the total covering factor of many clouds in a spherical layer from $R$
to $R+dR$ (the differential CF) is
\begin{equation}
CF=4\pi \Omega \ns R^2\sin \ief dR,
\end{equation}
where $\ns$ is the stellar density (the number of clouds per unit
volume is close to the stellar number density because we assume the life-time of a
cloud is close to half the orbital period of the star: one star -- one
cloud).
\begin{figure*}
 \parbox{\textwidth}{
   \parbox{0.3\textwidth}{\epsfxsize=0.3\textwidth\epsfbox{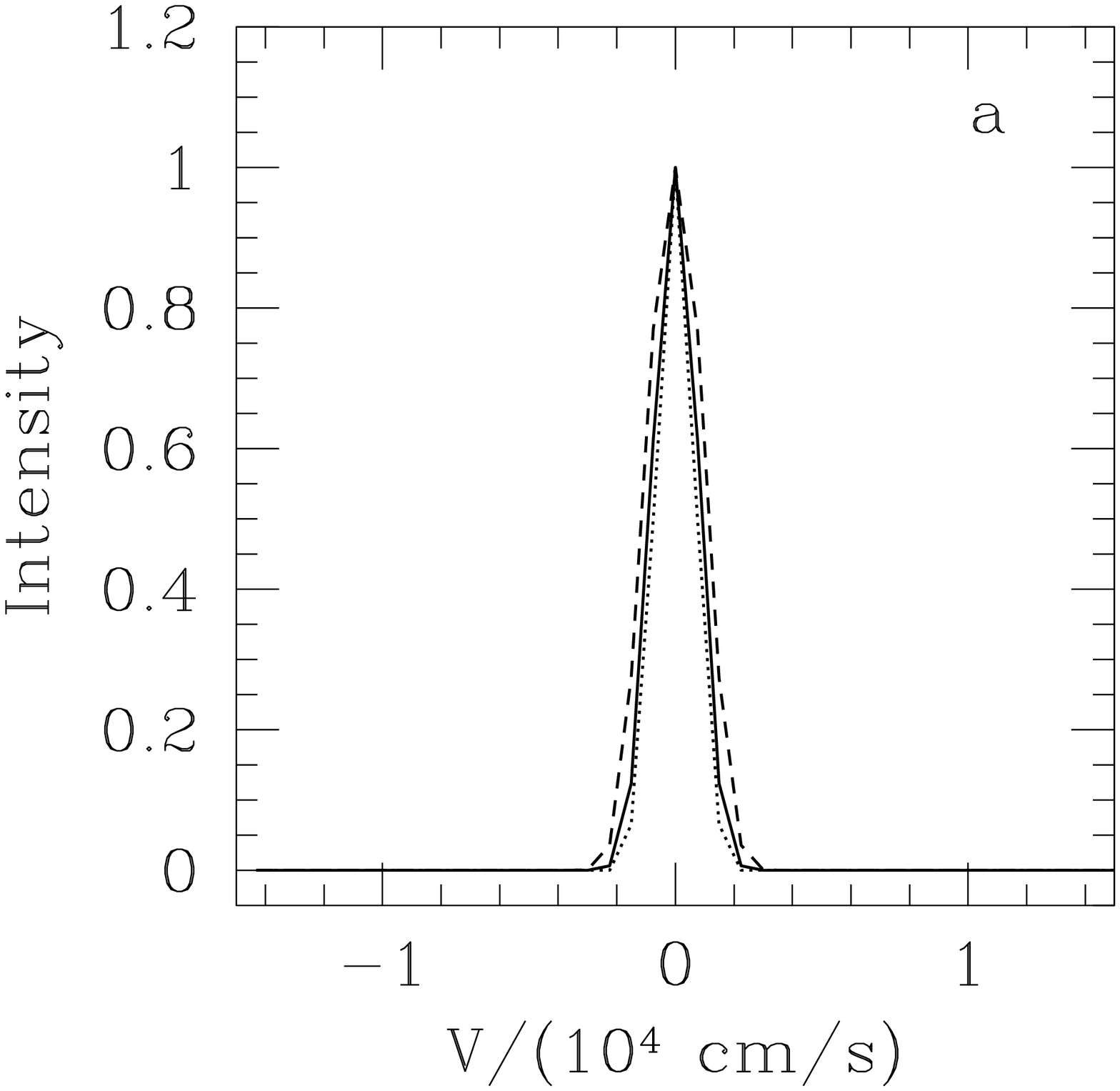}}
   \parbox{0.3\textwidth}{\epsfxsize=0.3\textwidth\epsfbox{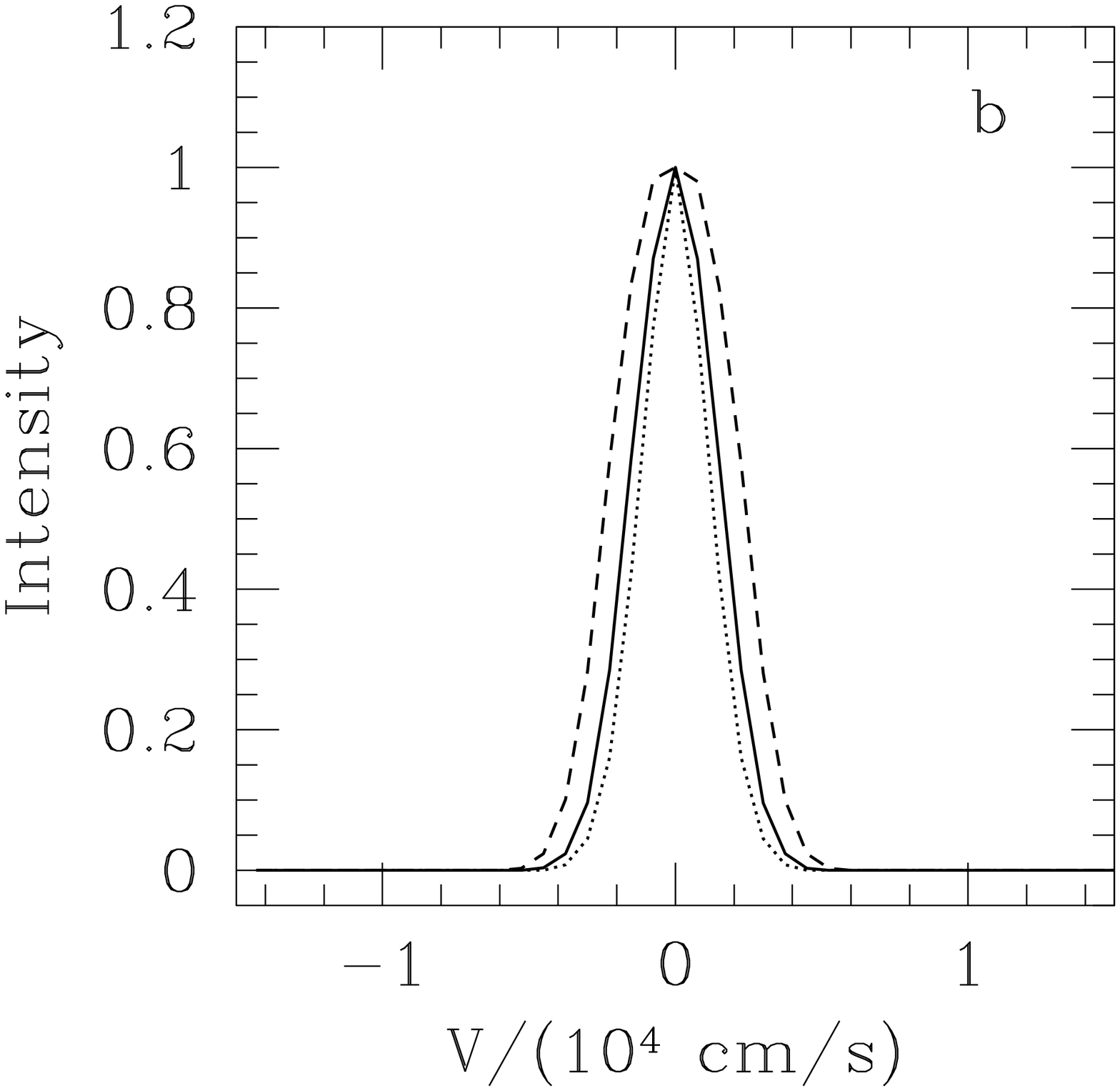}}
   \parbox{0.3\textwidth}{\epsfxsize=0.3\textwidth\epsfbox{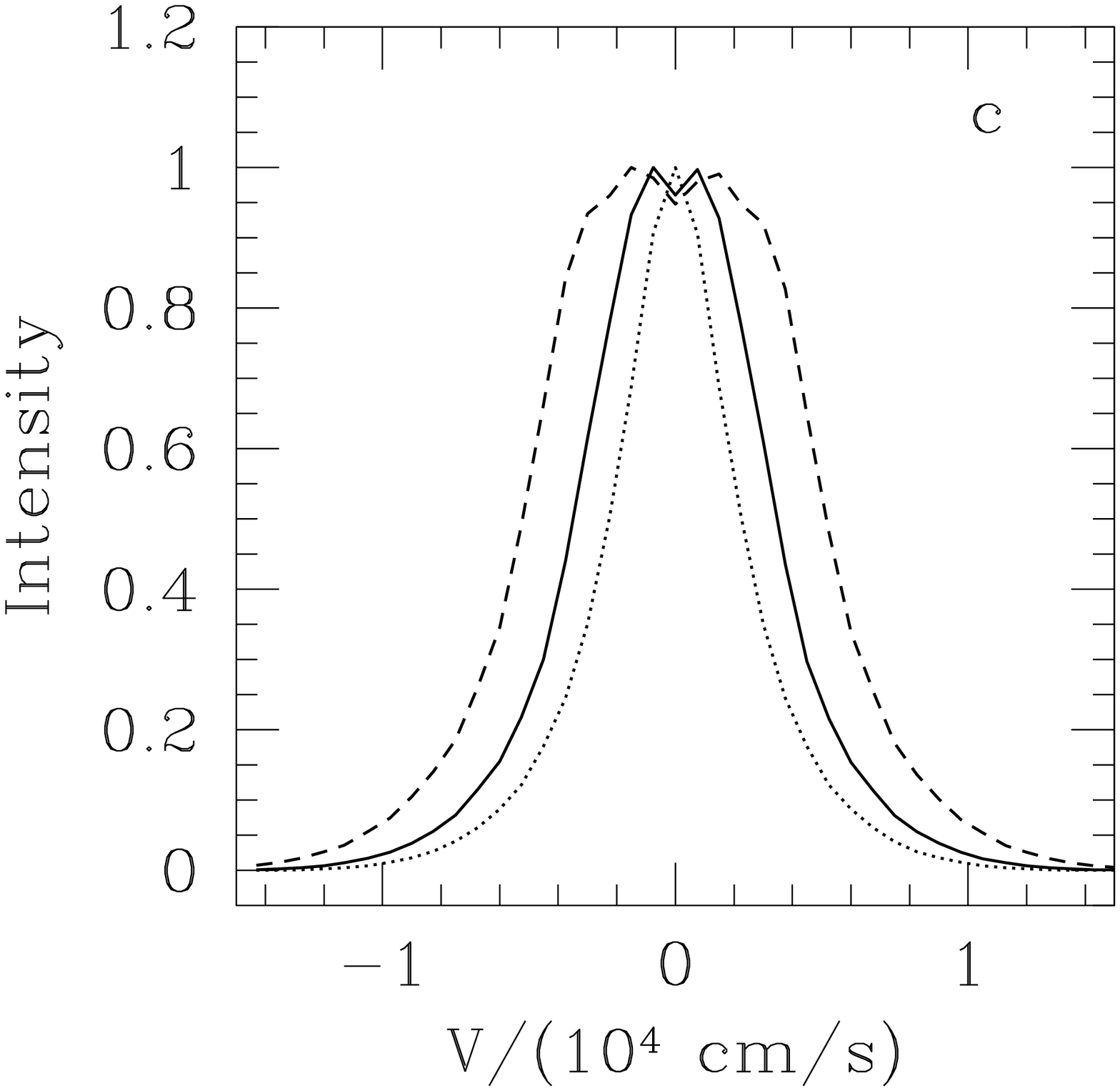}}
}
\caption{
$H_\alpha$ line profiles, calculated with $\MBH=10^7M_{\odot}$(a),
 $\MBH=10^8M_{\odot}$(b), and $\MBH=10^9M_{\odot}$(c). Dotted, solid and
dashed lines correspond to $\theta =10^{\circ },30^{\circ }$ and $60^{\circ }$
}
\end{figure*}
Taking into account that at small and intermediate inclination angles $\phi_
1/(\sin \ief)\sim \sin \ief$, and that the flow of the ionizing photons is
proportional to $\Phi _{\rm i}\sim 10^{22}m_8\dot mr_{-2}^{-2}$, 
one can see that
the particle densities and the radiation flows are within the range of
calculations of Korista et al. (1997) for the atlas of equivalent widths ($
EW $) of quasar broad emission lines. These calculations give $EW$ as a
function of two parameters, $\Phi _{\rm i}$ and $n$, supposing that all the clouds
have column densities $N=10^{23} ~{\rm cm}^{-2}$ and the total covering factor of the
clouds is $CF=1$. This permits us to calculate the profile of an emission
line as

\begin{eqnarray}
\lefteqn{
I(V)\simeq K\int_{R_{\min }}^{R_{\max }}EW(\Phi _{\rm i},n)\Omega \ns R^2\sin
\ief dR}\nonumber\\
& & \int \int \{C_{\rm r}(\varphi ,\theta )/(v_{\rm r}\cos \varphi)
 \pi ^{-1/2}D^{-1}\exp[-(v_{\rm ch}/D)^2]\nonumber\\
& & \delta (V-v_{\rm p}-v_{\rm ch})\} d\varphi dv_{\rm ch}
\end{eqnarray}
where $V$ is the velocity projection along the line of sight, $K$ is a
constant for a specific line, $EW(\Phi _{\rm i},n)$ is the $EW$ of the line
calculated in Korista et al. (1997) as a function of $\Phi _{\rm i}$ 
and $n$, $
\Omega $ is the coverage factor of a cloud at the distance $R$, $\ns$ is the
number
density of stars (and so of the clouds), $C_{\rm r}$ is the relativistic 
correction, $
v_{\rm p}(\varphi ,\theta )=v_{\rm r}\cos \varphi \sin \theta $ is the projection 
of the
rotational velocity $v_{\rm r}$ onto the line of sight, ($\varphi $ 
is the azimuth
angle and $\theta $ is the angle between the line of sight and the axis of
the disk), $v_{\rm ch}$ is the projection component of the chaotic velocity of
stars with dispersion $D(R)=0.67\cdot 10^9(M_8/r_2)\sin \ief$ cm s$^{-1}$, 
where $
M(R)$ is the total mass within radius $R$ and $\delta ()$ is the
delta-function.

Note that the velocity dispersion $D$ of the chaotic motion contains $%
\sin \ief$, because the velocity dispersion is due to the distribution
of the inclination angles and eccentricities around the effective values;
we assume here that the distribution of the projection velocities is
Gaussian in the coordinate system co-rotating with the AD.

We convolve the rotational velocity with the chaotic velocities of the stars;
this means that the velocity of clouds contains both the rotational (disk) and
chaotic velocity components of the stars. $R_{\min }$ and $R_{\max }$ in the
integral limits are taken correspondingly as $0.001{\rm pc}$ and $(m_8\dot
m)^{1/2}{\rm pc}$, the last is approximately determined by the inner
radius of the putative obscuring torus.

We performed the computations for the disk surface density profile of NT 
given by Eq.~(\ref{eq:NT}).

The results of the numerical calculations of the profiles of the $H_\alpha $
line, normalized to $1$ at the maximum, are presented in Fig.~4. Each
panel contains three profiles, corresponding to different inclination
angles of the line of sight to the axis of the AD: $\theta =10^{\circ
},30^{\circ }$ and $60^{\circ }$. The general shapes of the profiles are
determined by the combination of a broad part, which depends mostly on the
rotation of the inner region of the disk, and a more narrow central part,
which is determined mostly by stellar chaotic motions in the outer parts of
the system.

In Fig.~4(a-c) we show line profiles for the Eddington mass inflow ($\dot
m=1 $) and for different masses of the MBH, $m_8=0.1,1,10$, which would
correspond to a
Narrow-Line Seyfert 1 galaxy (NLS1), a Seyfert galaxy and a quasar, respectively. 
Our
calculations support the idea that the narrow line Seyfert 1 galaxies
are those with small MBH mass and almost Eddington mass inflow; or they can be
objects with intermediate MBH mass and with a mass inflow less than the
Eddington one, seen at an angle $\theta <30^{\circ }$.

Our theoretical consideration of the stellar distribution
above the AD gives reasonable line profiles, supporting the idea that the
emitting clouds are somehow related to the stars. They can be the wakes,
following the stars after crossing the disk, or/and some gas envelopes of
the stars themselves, which are disturbed by disk-crossing and by close
passages of other stars. The mean distance between the stars at stellar
density $\ns\sim 10^9~{\rm pc}^{-3}$ is $dr\sim 10^{-3}$ pc, and many stars
pass at even smaller distances, exciting each other by tidal forces and
tangential collisions. However, in the present paper we used the specific
model of clouds as the stellar wakes.

In our model the covering factor of the clouds along the AD plane became
higher than $1$ at large radii, so the spectrum of ionizing radiation has to
be changed in the outer parts of the BELR. 
Of course, our model calculations
are preliminary and rough, and a more detailed model has to be considered
for the sake of comparisons with the observed profiles. But the common
property following from the physical sense of our approach, viz. that the wings
of the profiles should be more variable than the central peaks, seems to be
confirmed by reverberation investigations.

\section{Variability}

In the present section we will estimate the level of variability driven by
the stars. Obviously, the interpretation of the variability demands the
calculation of rather complicated models, but here we will make a simple
analysis of the variability expected from the inelastic processes of
star-disk and star-star interactions. We consider two types of processes,
connected with the energetic events of s-d and s-s collisions: flashes
of UV radiation and X-ray flares. We consider a very simple model: 
let every event produce a fireball, filled with matter dragged from
the disk (s-d crossing) or from stars (s-s collision), and expanding in a
hot coronal medium. When the fireball becomes Compton-transparent, it
produces a UV flare. The radiation of the flare interacts with the hot
electrons of the corona above the disk, thus producing a flare of the
up-scattered X-radiation.

The real picture is of course much more complicated. The fireball radiation
spectrum evolves; quickly expanding fireballs create blast waves in the
environment, and the bow shock can be accelerated when propagating from the
disk surface into the corona with diminishing density. It is possible also
that the matter of the fireball is heated later on to the coronal
temperature, thus creating a transient patchy corona. The undisturbed,
steady-state corona (without any influence of stars) can extend up to $%
R=300\Rg$ (Kawaguchi et al. 2001); the patchy corona
created by s-d interactions can extend to even larger distances.

All these processes require a more detailed description, but here we consider
the simplest one, based on the simple picture of fireballs.

\subsection{The star-disk events}

When a star comes out of the disk, it is preceded by a bow-shock,
accompanied by a fountain of the matter from the disk. The total amount of
gas thrown out above the disk consist of two parts: $\Delta M_{\rm d}$ is the mass
dragged out from disk by the star, and $\Delta \Ms$ is the mass lost by the star.

We assume that the fireball is created by the first mass, and that the second
mass creates the wake, following the star and showing up as a BELR cloud, as
was discussed in the previous section.

Then the mass of the fireball is 
$M_{\rm f}=\Delta M_{\rm d}=q_1\pi \Rs^2\Sigd$,
where $q_1\leq 1$; numerically
\begin{equation}
M_{\rm f}\simeq 1.6\cdot 10^{26}q_1\rs^2\Sigma _4 ~{\rm g}.
\end{equation}

Let the fireball contain the energy deposited by the star into the disk
matter,
\begin{equation}
\Delta E\simeq 0.7\cdot 10^{44}Q\rs^2\Sigma _4m_8r_{-2}^{-1}\phi (i)/(\cos i)
~{\rm erg}.
\end{equation}

We assume that the fireball expands and produces the flare when it
becomes transparent, that is optically thin relative to the electron
scattering. At this moment its radius fits the relation 
$R_{\rm f}\rho \sigma =1,$ where 
$\rho =M_{\rm f}/(4/3\pi R^3)$ is the density of the fireball, and $\sigma
\simeq 0.4~{\rm cm^2~g^{-1}}$ is the Thomson scattering cross section per gram of the
ionized cosmic plasma. Then the radius of the fireball at the moment of the
flare is
\begin{equation}
R_{\rm f}\simeq (M_{\rm f}\sigma _{\rm T}/4)^{1/2}\simeq .4\cdot 10^{13}(q_1\Sigma _4)^{1/2}\rs~
{\rm cm}.
\end{equation}
The density at the moment of the flare is
\begin{equation}
\rho _{\rm f}\simeq M_{\rm f}/(4/3\pi R_{\rm f}^3)\simeq 6.3\cdot
10^{-13}q_1^{-1/2}\rs^{-1}\Sigma _4^{-1/2},
\end{equation}
so the particle density is
\begin{equation}
n=3.2\cdot 10^{11}q_1^{-1/2}\rs^{-1}\Sigma _4^{-1/2}~{\rm cm}^{-3},
\end{equation}
and the column density is close to $N_{\rm H}=10^{24}~{\rm cm}^{-2}.$

The temperature in the fireball changes due to its expansion approximately
as
\begin{equation}
T_{\rm f}\simeq [q_1\Delta E/(4/3\pi aR^3)]^{1/4}~{\rm K},
\end{equation}
where $a=7.6\cdot 10^{-15}$ erg~cm$^{-3}$ K$^{-4}$, 
and $R$ is the radius of the
expanding fireball.

At the moment of the flare, the temperature is therefore
\begin{equation}
T_{\rm f}\simeq .8\cdot 10^5Q_{\rm T}\rs^{-1}\Sigma _4^{-1/2}m_8^{1/4}r_{-2}^{-1/4}\phi
1^{1/4}~{\rm K},
\end{equation}
where $Q_{\rm T}=q_1{}^{-3/8}Q^{1/4}$, and the spectral maximum of the flare is in
the far UV-band,
\begin{equation}
\nu \sim 5\cdot 10^{15}~{\rm Hz}.
\end{equation}

The estimate of the minimum duration of the flare (assuming that the
radiation leaves the fireball when the latter becomes Compton-thin) is
\begin{equation}
\Delta t_{\rm f}\leq 2R_{\rm f}/c\simeq .3\cdot 10^3q_1\Sigma _4^{1/2}\rs ~{\rm s},
\end{equation}
and the average luminosity during the flare is
\begin{equation}
L_{\rm f}=\Delta E_{\rm f}/\Delta t_{\rm f}\leq 3.\cdot 10^{41}q\rs\Sigma
_4^{1/2}m_8r_{-2}^{-1}\phi_1 ~{\rm erg~s}^{-1},
\end{equation}
where $q=q_1Q$.

In the NT disk
we have
\begin{equation}
L_{\rm f}\simeq 0.5\cdot 10^{43}q\rs m_8^{9/8}\alpha _{-2}^{-2/5}(\dot
m_1)^{7/20}r_{-2}^{-11/8}\phi_1 ~{\rm erg~s}^{-1}.
\end{equation}

The real duration of the flare is rather longer, and the luminosity is
less than in these estimates. When the temperature of the fireball becomes
lower than $10^5$ K, the plasma in the fireball partly recombines, which
increases the opacity to above its Thomson value. Besides, if the fireball expands
with a velocity close to the sound velocity $V_{\rm f}\simeq
(k_{\rm B}/m_{\rm p}T_{\rm f})^{1/2}\simeq 3\cdot 10^6$ cm s$^{-1}$, the time of expansion is $
t_{\rm ex}\simeq R_{\rm f}/V_{\rm f}\simeq 10^6$ s. So, rough estimates show that the
characteristic duration of the flare is in between the two
limits
\begin{equation}
3\cdot 10^2q_1\Sigma _4^{1/2}\rs<\Delta t_{\rm f}(\rm s)<10^6q_1\Sigma
_4^{1/2}\rs/(T_{\rm f5})^{1/2}.
\end{equation}

Now we estimate the combined effect of all star-disk events.

It has to be noted here that the results of model calculations of
luminosity and variability, connected with the star-disk interaction,
depend strongly on the values of the rates of energy change, estimated
by Eqs.~(\ref{eq:ss1}) and (\ref{eq:sd1}). In principle, these estimates
can be increased, because the coefficient Q in $d\Esd/dt$ is
unspecified, and $d\Ess/dt$ can be increased, e.g. due to collective
processes in the flat distribution of stars. The same effect of increasing
luminosity is expected if the equilibrium $d\Esd/dt=d\Ess/dt$ is not
reached, due to larger values of $\sin(\ief)$. Because of this, we
calculated all quantities of interest (see below) for our reference
models with $m_8=0.1,1,10$ for two cases: 1) energy change rates given by
Eqs.~(\ref{eq:ss1}) and (\ref{eq:sd1}); 2) energy change rates increased by a factor
of 5.

The mean luminosity provided by star-disk interaction in the ring between $R$
and $R+dR$ is $dL=\Delta E_{\rm f}d\nu _{\rm sd}$, with $d\nu _{\rm sd}$ from 
Eq.~(\ref{eq:nusd}):
\begin{eqnarray}
\lefteqn{ dL\simeq 2\cdot 10^{43}Q\Sigma _4[\dot
n_{\rm y}^r/(\ln\Lambda)_1]^{1/2} \times }  \nonumber \\
& & \rs^2\ms^{-1}m_8^{9/4}r_{-2}^{-11/4}\phi (i)\sin (\ief)dr_{-2},
\end{eqnarray}
and the total luminosity of the flares is
\begin{equation}
L_{\rm fT}=\int L_{\rm f}\Delta t_{\rm f}d\nu _{\rm sd}=\int \Delta E_{\rm f}d\nu _{\rm sd}.
\end{equation}

From calculations of the reference models with $m_8=0.1, 1, 10$, we have the
total UV-flare luminosities (in units of $10^{43}~{\rm erg~s}^{-1}$) in
the first case $L_{\rm fT}\simeq 0.02, 0.07, 0.39$, and for the second case ($
dE/dt $ 5 times larger): $L_{\rm fT}\simeq 2.2, 29, 7.4$.

One can see that in the second case the total luminosity of flares resulting
in the s-d interactions is comparable to the variable fraction of the UV
luminosity.

The average number of the simultaneous flares in every $dr$
--ring is $dL/L_{\rm f},$ so the total number of coexisting flares is
\begin{equation}
N=\int \Delta t_{\rm f}d\nu _{\rm sd}.
\end{equation}
For the three MBH masses ($m_8=0.1, 1, 10$), we have (with $\Delta t^{min}$) in
the first case $N\simeq 3.6, 5.7, 100$, and in the second case $N\simeq
17, 73, 300$.

The weighted luminosity of a flare is $\tilde F_{\rm f}=L_{\rm fT}/N$, and the
mean-square deviation of the luminosity due to chaotic variability is
\begin{equation}
DL_{\rm sd}=\tilde F_{\rm f}N^{1/2}=L_{\rm fT}/N^{1/2}.
\end{equation}
Corresponding values for the three MBH masses are in the first case $%
DL/L_{\rm f}=1/N^{1/2}\simeq 0.5, 0.4, 0.1$, and in the second case $DL/L_{\rm f}\simeq
0.24, 0.12, 0.06$; so, the variability amplitude diminishes with increasing
luminosity of AGNs.

\subsection{The star-star events}

Similarly we can estimate the variability connected with the star-star
contact collisions. We assume that the average mass-loss per collision is \\
$M_{\rm f}\sim 0.1\zeta _{-1}\Ms=2\cdot 10^{32}\zeta _{-1}\ms~{\rm g}$.\\

The parameters of the s-s fireballs (the total energy, the radius of the
fireball, the temperature of UV radiation, the minimum and maximum duration
of the UV flare are:
\begin{eqnarray}
\Delta E & = & 0.45\cdot 10^{50}\zeta _{-1}\ms m_8r_{-2}^{-1}\sin ^2(\ief)~{\rm erg}, \nonumber \\
R_{\rm f} & \simeq & (M_{\rm f}\sigma  _{\rm T}/4)^{1/2}\simeq 1.1\cdot 10^{16}\ms^{1/2}\zeta
_{-1}^{1/2}~{\rm cm}, \nonumber \\
T_{\rm f} &\simeq & 3.6\cdot 10^4\zeta _{-1}^{1/4}\ms^{-1/4}m_8^{1/4}r^{-1/4}\sin
^2\ief~{\rm K},\\
 \Delta t_{\rm f}^{\min}&\geq &2R_{\rm f}/c = 0.7\cdot 10^6\ms^{1/2}\zeta _{-1}^{1/2}~{\rm s} \nonumber, \\
\Delta t_{\rm f}^{\max}&\leq &2R_{\rm f}/V_{\rm sound}\simeq 3\cdot 10^9~{\rm s}. \nonumber
\end{eqnarray}

The corresponding average luminosity is given by
\begin{equation}
L_{\rm f}^{\max}\leq .64\cdot 10^{44}\zeta _{-1}^{1/2}\ms^{1/2}m_8r^{-1}\sin
^2\ief~{\rm erg~s}^{-1},
\end{equation}
\begin{equation}
L_{\rm f}^{\min}\geq 2\cdot 10^{41}\zeta _{-1}^{1/2}\ms^{1/2}m_8r^{-1}\sin ^2\ief
{\rm ergs}^{-1}.
\end{equation}

By definition, the column density of the fireball at the moment of the flare
is $N=Rn\sim 10^{24}~{\rm cm}^{-3}$, so $n\simeq 10^9~{\rm cm}^{-3}$.

Taking $d\nu _{\rm ss}$ from Eq.~(\ref{eq:dnss}), we have a number of 
simultaneous flares $N=\int \Delta t_{\rm f}d\nu _{\rm ss}$, which is typically less
than 1. This means that s-s events lead to even lower average luminosity
than s-d ones, though the amplitude of the s-s flares are larger.

\subsection{The relation of the UV and X-ray events and the frequency dependence}

Of course, the variability connected with stars may not be the only
mechanism operating in AGN. Their variability can be mostly driven by the
variable (unstable) accretion rate onto the MBH. The process of accretion
itself is rather complicated, and the theory develops quickly (see the
recent review in Park \& Ostriker, 2001). The variability observations show
some common properties of the variability of the AGNs and the stellar-mass
Galactic black hole candidates (Utley \& McHardy 2001, Chiang et all. 2000), 
and for that class of objects
the model of distributed flares has been considered (Galeev et al. 1979, 
Czerny \& Lehto 1997, Poutanen \& Fabian 1999). On the other hand, 
it has been shown that
UV and X-ray variability of AGNs cannot be explained with models assuming
simple reprocessing of the X-ray emission of a variable point source 
placed at the symmetry axis above AD. 
Is it possible that s-d collisions can work in AGNs instead of
the magnetic flares supposed to exist in the Galactic black holes?

From the rough estimates provided in the previous subsection one can
conclude that the s-d interaction can lead to a noticeable variability
amplitude only if both rates of energy dissipation -- due to s-d and s-s
interactions -- are enhanced several times in comparison to our basic
estimates. This can possibly be achieved with a more realistic distribution of
stellar masses. An increase might also come due to unusual properties
of the stars in the very compact stellar clusters. Deviations from the
equilibrium conditions (which was assumed by us for the approximate
solutions) also promote larger values of $\sin(\ief)$ and, as a
consequence, lead to more power radiated from s-d interactions.

Obviously, the UV events are divided into two types: those connected with
s-d and with s-s interactions. The relations of the two types of interactions 
to the X-ray events are different.

In the s-d case, the UV flares occur at relatively small heights above the AD,
which are close to the fireball size, $h\sim 0.39\cdot 10^{13}(q_1\Sigma
_4)^{1/2}\rs$ cm. The time for reaching this distance is 
$t\sim R_{\rm f}/V_{\rm f}$,
where $R_{\rm f}$ and $V_{\rm f}$ are the size of the fireball and 
the velocity of its
expansion, and the time from the appearance of the blast wave to the maximum of the
UV luminosity is $t\sim 0.39\cdot 10^4(q_1\Sigma _4)^{1/2}\rs$ s. The size of
the region of up-scattering is about the coronal scale height,
around $10^{13}-10^{14}$ cm, which leads to a time-lag between UV and
X-ray flares of about $\Delta t_{\rm f}\sim 10^3-10^4$ s.

In the s-s case, most events occur at large heights above
the disk, so the fraction of photons reflected by the corona is small, and
the duration of Compton-scattered X-ray events, as well as the time-lag, are
much longer than in the case of the s-d event.

Using the estimates from the previous section, we can consider the
amplitude-frequency relations following from star-disk and star-star
dissipative interactions. In Fig.~5 we plot the dependences of the UV flare
luminosity $L_{\rm f}=\Delta E_{\rm f}(R)/\Delta t_{\rm f}(R)$, 
divided by frequency $d\nu (R)$
on the frequency. We plot both $L_{\rm fss}/d\nu _{\rm ss}$ 
and $L_{\rm fsd}/d\nu _{\rm sd}$
in the same figure using the above-defined dependence of both values
on radius and replacing the radial coordinate with the frequency coordinate
where the frequency-radius relation is taken from the 
determined characteristic 
frequency of events at a given radius.
Let us call the dependence
``frequency-amplitude dependence'' (FAD). Note that the FAD cannot be
strictly compared to the power density spectrum (PDS), observed in the UV
and X-ray bands (e.g. McHardy \& Czerny 1987, Chiang et all. 2000). 
These PDS are obtained by
Fourier transformation of the light curves, which are usually measured 
in frequency bands about $10^{-7}-10^{-4}$ Hz and show a sort of red
noise, that is $PDS\propto \nu ^{\rm -p}$ with $p\sim 1-2$; sometimes 
$p\sim 0$ is found at the low-frequency limit. The FAD spectrum shown in 
Fig.~5 has a different meaning: it is just the dependence of the flare
luminosity per frequency interval on the frequency of the events. Besides,
we recall that we use the equal stellar mass and other approximations,
so the real slopes of the curves could be steeper and the curves 
could cover a broader
frequency range. Nevertheless,
some general prediction about PDS can be derived from our calculations:
\begin{figure}
 \epsfxsize=0.45\textwidth \epsfbox{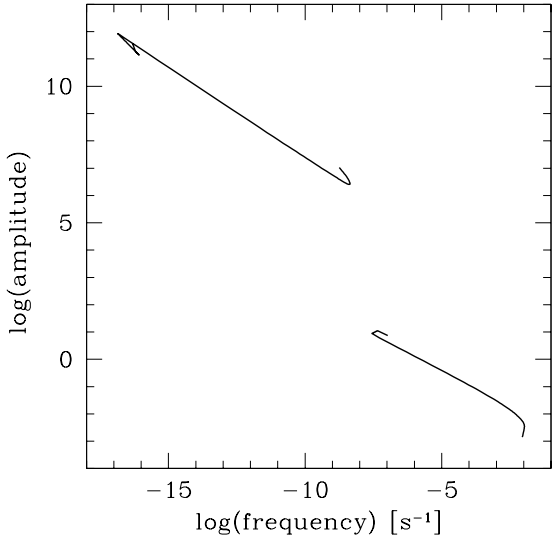}
 \caption{The Frequency-Amplitude dependence, including s-d (high-frequency)
and s-s (low frequency) interactions }
 \end{figure}

i) the PDS of AGNs can be divided into two parts: fast variability (on
time scales from hundreds of seconds to about a year, which is
connected with the s-d processes) and slow variability (on time scales 
from several years to thousands of years, connected with the s-s
collisions).

ii) the PDS slope can change abruptly somewhere in between the fast and
slow part, and the slow variability part of the PDS has a larger amplitude.
This conclusion seems to be in agreement with the results of long-time
observations in X-ray band (Markowitz \& Edelson 2001), and it is
definitely supported by long-time observations in the optical band
(Czerny et al. 1999, Czerny et al. 2001).

iii) the differences of PDS between the slow and fast parts may depend on
the AGN luminosity and other parameters.

Our preliminary conclusions are that the presence of a CSC in AGNs does not
contradict the observations, the CSC can contribute to AGN variability in some
cases, and further comparison of variability of AGNs and Galactic black
holes is needed.

\section{Discussion and conclusions}

The role of the compact stellar cluster in the creation and evolution of AGN
is still poorly understood. We suppose that this role is very important, so
that the AGN phenomenon can be considered as a result of interactions of
three main physical subsystems: the CSC, MBH and gas subsystem. We call this
approach the interacting subsystems theory (Vilkoviskij et al. 1999).

Our model of the CSC-disk interaction, though very approximate, is the first
one which treats star-star and star-disk interactions jointly, so it can
serve as a starting point for considering some new physics in AGN problems.

We showed that the star-disk interaction tends to order stellar orbits,
dragging them to the disk plane and making them circular, but the star-star
gravitational interaction scatters these orbits and prevents them from being
incorporated into the disk. We assume equilibrium of the two processes
and use a simple inflow model instead of analyzing the kinetic equations.
The other simplification was the assumption of the equal stellar masses,
which is acceptable as a first step only.

Our results can be compared with the results of Rauch (1995, 1999). In the
last work he calculated numerically the evolution of the
spherically-symmetric stellar cluster with stellar collisions, but without
s-d interactions. He finds that the internal stellar density distribution
achieves a quasi-stationary state
in the late stages. Our calculations of the
radial stellar density distribution with direct stellar collisions and s-d
interactions are in qualitative agreement with those of Rauch (1999).

On the other hand, our calculations of the shape of the stellar density
distributions in the central parts of AGNs, are naturally different to
the results of Rauch (1995), where he considered the evolution of
individual stellar orbits crossing the disk (without s-s interaction). He
calculated the evolution of the stellar system, which surrounds the AD in
the first moment, and showed that the evolution leads to the completely flat
system, incorporated into the disk plane in the final stage. In contrast, we
take into account (besides the s-d interaction) both elastic and non-elastic
star-star interactions. This leads to a quasi-stationary distribution of
stars, which is flat only in the central part of the disk.

The determination of a more realistic stellar distribution in the innermost
part of the central cluster allowed us to consider some consequences of the
complex system of AGN, containing MBH, AD and CSC.

We showed that the simple model of stellar wakes accompanying the s-d
interactions gives profiles of broad emission lines similar to those observed,
providing a reasonable solution for the puzzles of both generation and
dynamics of the BELR clouds. We used even further simplified models for
estimates of the variability connected with s-d and s-s interactions, and
we come to the preliminary conclusion that a noticeable role of the stars in
the variability is possible, but only if both dissipative and scattering
interactions of stars are enhanced relative to our preliminary estimates.
The variability produced by stars is naturally divided into a fast and
a slow parts, corresponding to s-d and s-s processes.

Of course, the main questions remains whether the stellar density of CSC in
AGNs is large enough for all the effects discussed to proceed efficiently.
Here we assumed that the masses of CSC in the bright AGNs are about an order
of magnitude larger than the MBH mass.

The real stellar density and masses of the CSC in AGNs are still unknown. In
the paper of Schinnerer et al. (2001) the near-infrared high
angular resolution speckle imaging of the central region of the Seyfert
galaxy NGC 3227 was performed. The authors showed that the CSC do exist in
the inner $300$ pc, and that in the innermost (unresolved, $<70$ pc) part
there is the even more dense core of the stellar cluster, which contributes
less to the luminosity, but contains 85\% of the mass of the cluster. This
result supports our leading idea of large enough masses of CSCs in AGNs. A
fuller investigation of 112 Seyfert galaxies in $1.6\mu$m with the
Hubble Space Telescope was performed by Quillen et al. (2001). They showed
that there are unresolved sources in all Seyfert galaxies with
luminosities $L(1.6 ~\mu {\rm m})\sim (10^{41}-10^{42})~{\rm erg~s}^{-1}$ 
correlating
with O[III] $\lambda 5007$ and with hard X-ray luminosities for Seyfert 1.0-1.9
sources. The luminosities are much lower in Seyfert 2.0 sources. 
Due to this trends
the authors concluded that the unresolved sources are mostly of non-stellar
(hot dust) origin; but we note that in our model the correlation of the
AGN luminosity with the mass of the CSC is expected. More detailed
observations are needed for solving the question about CSC masses in
AGNs.

It follows from our models that the accretion disk itself can be created as
the result of mass loss due to contact stellar collisions in the rotating,
compact and massive CSC. In this case the collision disruptions of stars can
partly feed the accretion disk flow, and another part of this gas (the hot
part) can support the outflow of the hot gas from AGNs, as was supposed by
Vilkoviskij et al. (1999). This opens new possibilities for the more
self-consistent models of the AGN structure, where the accretion disk,
obscuring torus and polar outflow are created (partly at least) due to
stellar collisions in the rotating CSC. The transfer of angular momentum
from AD to the CSC also has to be taken into account.

One can imagine the following general picture of the AGN's duty cycle:

The bright phase of AGNs should be relatively short, about several times $10^8$
years, and usually the activity cycle consists of three phases. It starts
with creating a new CSC in the dense gas cocoon around the MBH (say,
generated by a galactic merger), which corresponds to the IR-bright {\sl IRAS}
galaxies (the first phase). In this phase a new CSC is formed due to the
powerful star-burst in the galactic center, which leads to the 
creation of a new
AD, thus initiating the star-disk dynamics, similar to what was considered by
Rauch (1995).

The subsequent growth of the AD luminosity, the creation of the central flat
stellar distribution, and the appearance of the polar jets and the hot-gas
outflows make polar holes in the cocoon, and the second
phase of activity sets in; the AGN looks like a 
typical QSO or Seyfert galaxy with an
obscuring torus. Now in this phase most of the stars in the polar regions of
the rotating CSC are exhausted, and stars cross mostly the periphery of the
AD due to the diffusion into an effective loss-cone. The AD itself can
be supported by stellar collisions in the rotating CSC, and our solutions
approximate stellar dynamics in this phase. During approximately $10^8$
years the mass of the CSC and the stellar inflow will gradually decrease, thus
leading to the third phase with diminished activity, corresponding to the
weak Seyferts and LINERs (galaxies with Low Ionization Narrow 
Emission Region).

Therefore, the results of our paper correspond to the second phase, the
`adult and strong' age of an AGN.

It is clear that more detailed model calculations are required. Our
description of star-star interaction in the non-spherical, flat stellar
system is very rough and the results should be verified with n-body
simulations. Advanced models should also take into account the distribution
of stellar masses and a more accurate description of the accretion disk. New
achievements in theory and observations of AGN show that the
accretion flow is possibly a combination of a geometrically thin cold disk
in the outer part, and an advection-dominated accretion flow (ADAF) 
in the central part of the disk in some cases
(Ichimaru 1977; Rees et al. 1982; Narayan \& Yee 1994; Abramowicz et al.
1995, Loska \& Czerny 1997), with the hot corona still present above the
cold disk (Esin et al. 1997; Esin et al. 1998, Witt et al. 1997).

However, even the approximate results presented here show that compact and
massive stellar clusters in the centers of AGNs can play an important role
in the creation and physical properties of the AGN engine.

\begin{acknowledgements} E.Y.V. is grateful to Suzy Collin for
many  discussions, initiating this work, and to CAMK for the kind
hospitality permitting to write this work.

The work was supported by grant N 2P03D01816 of the Polish State Committee
for Scientific Research, and partly by the INTAS grant No 96-0328.
\end{acknowledgements}

\end{document}